\documentclass[showpacs,superscriptaddress,preprintnumbers,amsmath,
amssymb,nofootinbib]{revtex4}

\usepackage{amsmath,amssymb,latexsym}
\usepackage{color}
\usepackage{graphicx,bm}
\usepackage{dcolumn}    

\definecolor{red}{rgb}{1,0,0}
\definecolor{blue}{rgb}{0,0,1}
\definecolor{green}{rgb}{0,1,0}

\begin{document}

\newcommand{\ogw}{ {\Omega_{\mathrm{gw}}}  }

\title{ 
Detecting a gravitational-wave background with 
next-generation space interferometers 
}

\date{ }

\author{Hideaki Kudoh} 
\email{kudoh_at_utap.phys.s.u-tokyo.ac.jp}
\affiliation{Department of Physics, The University of Tokyo, Tokyo 113-0033, Japan }

\author{Atsushi Taruya} 
\email{ataruya_at_utap.phys.s.u-tokyo.ac.jp}
\affiliation{ Research Center for the Early Universe~(RESCEU), School
  of Science, The University of Tokyo, Tokyo 113-0033, Japan }

\author{Takashi Hiramatsu} 
\email{hiramatsu_at_utap.phys.s.u-tokyo.ac.jp}
\affiliation{Department of Physics, The University of Tokyo, Tokyo 113-0033, Japan }

\author{Yoshiaki Himemoto}
\email{himemoto_at_utap.phys.s.u-tokyo.ac.jp}
\affiliation{Department of Physics, The University of Tokyo, Tokyo 113-0033, Japan }

\preprint{gr-qc/0511145, UTAP-544, RESCEU-37/05}


\begin{abstract}  
Future missions of gravitational-wave astronomy will be operated by space-based interferometers, covering very wide range of frequency. 
Search for stochastic gravitational-wave backgrounds (GWBs) is one of the main targets for such missions, and we here discuss the prospects for direct measurement of isotropic and anisotropic components of (primordial) GWBs around the frequency $0.1-10$ Hz. 
After extending the theoretical basis for correlation analysis, we evaluate the sensitivity and the signal-to-noise ratio for the proposed future space interferometer missions, like Big-Bang Observer (BBO), Deci-Hertz Interferometer Gravitational-wave Observer (DECIGO) and recently proposed Fabry-Perot type DECIGO. 
The astrophysical foregrounds which are expected at low frequency may be a big obstacle and significantly reduce the signal-to-noise ratio of GWBs. 
As a result, minimum detectable amplitude may reach $h^2 \ogw = 10^{-15} \sim 10^{-16}$, as long as foreground point sources are properly subtracted. 
Based on correlation analysis, we also discuss measurement of anisotropies of GWBs.
As an example, the sensitivity level required for detecting the dipole moment of GWB induced by the proper motion of our local system is closely examined.
\end{abstract}

\maketitle

\section{Introduction} 
\label{sec:intro}      

The operation, construction and new projects of a number of new and next-generation gravitational-wave detectors are currently underway, and they will constitute a global network of detectors in a near future. 
Furthermore, future missions of gravitational-wave astronomy will be operated by space-based interferometers. 
As a result, gravitational-wave searches will be performed in a very wide range of frequency band. 
The current searches of gravitational waves (GWs) are mainly classified into four types; coalescing binary systems (e.g.,
\cite{Ando:2001ej,Abbott:2005pe,Abbott:2003pj}), continuous waves with very slow evolution \cite{unknown:2005pu}, (stochastic) gravitational-wave 
backgrounds (GWBs) \cite{Abbott:2005ez} and gravitational-wave bursts 
\cite{Ando:2004rr,Abbott:2004rt}. 
The subject of this paper is a stochastic background probed by future missions of space-based interferometers.

A stochastic background of gravitational waves could result from random superposition of an extremely large number of weak, independent and unresolved gravitational waves (sources). 
This type of GWB is produced in many processes during cosmological and astrophysical evolutions, so that the spectrum is characterized by the generation mechanism. 
For example, a standard inflation model predicts a nearly scale-invariant spectrum of 
$\ogw(f)$, where $\ogw(f)$ denotes the gravitational-wave energy density divided by the critical energy density to close the universe, whereas the large population of Galactic and extra Galactic binary systems makes up stochastic backgrounds known as a confusion noise, which sometimes dominates instrumental noises of a detector. 
Therefore, exploring GWBs brings us a new and interesting window to probe the early universe, as well as the astrophysical objects 
(see, e.g., Refs.~\cite{Enoki:2004ew,Koushiappas:2005qz,Sesana:2004sp,Sesana:2004gf,Smith:2005mm,Wyithe:2002ep}).

Currently, several future missions of space-based interferometers have been proposed as follow-on missions of Laser Interferometer Space Antenna (LISA)\cite{Bender:1998}.
A main target of these missions is the primordial GWB produced during the inflationary epoch. 
While the conceptual designs of these projects differ from each other, it is commonly believed that the frequencies around 
$0.1\lesssim f \lesssim 10$ Hz may be one of the best observational window filling the gap between the frequency covered by LISA and the ground-based detectors. 
We then wish to know the sensitivity of the next-generation space interferometers to the stochastic GWBs and to study the basic aspects of the signal processing strategy as well as the characteristics of each detector. 
To address these issues, the correlation analysis plays a key role since it is necessary to detect the stochastic signals in the presence of random noises. 
As we have mentioned, not only the instrumental noises but also the stochastic signals themselves become a disturbance and prevent us from the detection of primordial GWBs. In this respect, the optimally filtered signal-processing is suitable for the correlation analysis. 
In this paper, together with some extensions of the theoretical basis for the optimal-filtered signal processing, we discuss the sensitivity of the next-generation space interferometers and study the prospects for direct measurement of primordial GWBs.   
In addition, we also address the feasibility of direct measurement for anisotropic component of GWB.
As for future space interferometers, we will consider DECIGO/BBO \cite{Seto:2001qf,PhinneyBBO:2003,BBO:2003} and recently proposed Fabry-Perot type space interferometer (FP-DECIGO) (see \cite{FP-DECIGO:2005} for its pre-conceptual design).

The paper is organized as follows. 
In Sec. \ref{sec:preliminaries}, we begin by describing the properties of a stochastic background and statistical assumptions. In Sec.~\ref{sec:optimal_filter}, we analyze the signal processing required for the optimal detection of GWBs by space-interferometers, in both the isotropic  and the anisotropic GWB cases. 
Based on this, in Sec.~\ref{sec:sensitivity}, we study the sensitivity of next-generation space interferometers to the stochastic GWBs. 
We will quantify the minimum detectable amplitude of $\ogw$, taking account of the astrophysical foregrounds in the low frequency band. 
As an application of our formulation, we also discuss the detectability of dipole moment induced by the proper motion of our local system. 
Finally, Sec.~\ref{sec:summary} is devoted to a summary and discussion. 
Throughout the paper, we adopt the unit $c=1$.

\section{Preliminaries}
\label{sec:preliminaries}

Let us start by briefly reviewing basic concepts of data analysis for detecting stochastic GWBs. For readers familiar with these subjects, one may skip this section and move to Sec.~\ref{sec:optimal_filter}.

The gravitational-wave detectors measure the time variation of spacetime metric as one-dimensional time-series data. Denoting the signal strain measured by the interferometer $I$ (whose position is located at 
$\mathbf{x}_{\scriptscriptstyle I}$) by $h_{\scriptscriptstyle I}(t)$, it is expressed in terms of the superposition of the plane-waves by 
\begin{eqnarray}
 h_{\scriptscriptstyle I}(t) &=&
     \sum_{A=+,\times} \,\,  \int_{-\infty}^{ \infty} df  \,
\int_{S^2} d {\bf{\Omega}} \,\,
    \, {\mathrm{D}}_{\scriptscriptstyle I}^{ij}({\bf \Omega},f; t)\,\, 
    {\mathrm{e}}^A_{ij}( {\bf \Omega}) \, \widetilde{h}_A(f, {\bf \Omega})\,\,
    e^{ 2\pi i f (t-{\bf{\Omega  \cdot  x}}_{\scriptscriptstyle I})}, 
\label{eq:h_I}
\end{eqnarray}
where ${\mathrm{e}}^A_{ij}( {\bf \Omega})$ is the spin-2 polarization tensors and $\widetilde{h}_A(f, {\bf \Omega})$ is amplitude of gravitational wave. 
The quantity ${\mathrm{D}}_{\scriptscriptstyle I}^{ij}$ denotes detector tensor, which manifestly depends on an observed frequency $f$ and an arrival direction of gravitational waves. 
Further, it may also vary in time due to non-stationarity of detector's (orbital) motion. A functional form of the detector tensor generally depends on design of interferometer as well as a signal processing method and it plays a crucial role when discussing detectability of gravitational waves (Sec.~\ref{sec:sensitivity}).

By definition, the amplitude and polarization of GWB are statistically random, and the signal is usually assumed to be Gaussian with zero-mean.  
In this case, the statistical properties are completely characterized by the power spectral density $S_h $: 
\begin{eqnarray}
 \left\langle 
 \widetilde{h}_A^*  (f,  {\bf{\Omega}})
 {{\widetilde{h}}_{A'} }  (f', {\bf{\Omega}}')
\right\rangle
&=&
 \frac{1}{2}\, \delta(f-f')\, 
\frac{ \delta^2 (  {\bf{\Omega}},  {\bf{\Omega}}')}{4\pi}\,
 \delta_{AA'}\, S_h(|f|, \,\mathbf{\Omega}), 
\label{eq:def_GWBspec}
\end{eqnarray}
where $\langle~\rangle$ stands for an ensemble average. 
We do not assume isotropy of GWBs and hence the spectral density 
$S_h$ becomes the function of two-sphere $\mathbf{\Omega}$ in addition to $f$ \cite{Kudoh:2004he,Taruya:2005yf}. The power spectrum density $S_h$ is related to the dimensionless quantity $\ogw$ commonly used in the literature, which is the ratio of GW energy density $d \tilde \rho_{gw}$ contained in the frequency range $f$ to $f+df$ to the critical energy density, 
$\rho_{\mathrm{crit}}=3H_0^2/8\pi G$. 
Using Eq. (\ref{eq:def_GWBspec}) and the plane wave expansion 
(\ref{eq:h_I}) of the gravitational waves, one gets for $f \ge 0$ 
\begin{eqnarray}
    \ogw (f)  
    = \frac{1}{\rho_{\mathrm{crit}}} 
    \frac{d \tilde{\rho}_{\mathrm gw}}{d \ln f} 
    =
     \frac{8\pi^2}{3} \frac{f^3}{H_0^2} 
    \int \frac{d\Omega}{4\pi} \, 
    \frac{S_h (f, \Omega)}{2}, 
\label{eq:def_Omega_gw}
\end{eqnarray}
where the factor $\frac{1}{2}$ comes from our one-sided normalization in 
Eq.~(\ref{eq:def_GWBspec}).

An output $s_{\scriptscriptstyle I}(t)$ of detector is given by a superposition of gravitational-wave signal $h_{\scriptscriptstyle}(t)$ and instrumental noise $n_{\scriptscriptstyle I}(t)$, i.e.,
$ s_{\scriptscriptstyle I}(t) = 
   h_{\scriptscriptstyle I}(t) + n_{\scriptscriptstyle I}(t) $. 
Here, we assume that the noise obeys stationary Gaussian process. The power spectral density of noise is given by in Fourier space 
\begin{equation}
\langle \tilde{n}_{\scriptscriptstyle I}^*(f)
\tilde{n}_{\scriptscriptstyle J}(f')\rangle = \frac{1}{2}\,
\delta_{\scriptscriptstyle IJ}\,\delta(f-f')\,N_{\scriptscriptstyle I}(f), 
\label{eq:def_noise}
\end{equation}
In the presence of additive noises, a single-detector measurement cannot separate the signal $h_{\scriptscriptstyle}(t)$ from the instrumental noise and a reliable detection of the GWBs is impossible unless the amplitude of signal is large compared to the noises. 
Thus, one needs the other outputs and performs a correlation analysis. 
Provided several independent output data, the correlation analysis is examined by forming a product by multiplying data sets together and integrating over time:
\begin{eqnarray}
    S_{\scriptscriptstyle IJ}(t)
&\equiv&
    \int^{t+ \tau/2}_{t - \tau/2} dt'  \int^{t+ \tau/2}_{t - \tau/2} dt''  
    s_{\scriptscriptstyle I}(t') 
        s_{\scriptscriptstyle J}(t'') Q(t'-t''), 
\label{eq:correlation}
\end{eqnarray} 
where $Q(t)$ is an optimal filter function, which we will discuss in detail. 
$\tau$ is the local observation time corresponding to the time interval of chunk data. We assume that $\tau$ is smaller than the total observation time $T_{\rm obs}$, which is comparable to the time-scale of orbital motion of gravitational-wave detector.

Provided the product signal (\ref{eq:correlation}), detectability of the 
GWBs is quantified by defining the signal-to-noise ratio (SNR):  
\begin{eqnarray}
    \mathrm{SNR} = \left(\frac{T_{\rm obs}}{\tau}\right)^{1/2}\,\,
     \frac{\mu_{\scriptscriptstyle IJ}}{\Sigma_{\scriptscriptstyle IJ} }
     \equiv \left(\frac{T_{\rm obs}}{\tau}\right)^{1/2}\,\,
\frac{\left\langle S_{\scriptscriptstyle IJ}\right\rangle}{  
    \sqrt{
        \left\langle S_{\scriptscriptstyle IJ}^2 \right\rangle 
     -  \left\langle S_{\scriptscriptstyle IJ}\right\rangle ^2 }
     }.
\label{eq:def_SNR} 
\end{eqnarray}
In the above expression, the mean value 
$\mu_{\scriptscriptstyle IJ} = \left\langle S_{IJ} \right\rangle$ characterizes the GWB signal, and the noise contribution to it becomes vanishing when the noises $n_{\scriptscriptstyle I}$ and
$n_{\scriptscriptstyle J}$ are statistically independent of one another. 
On the other hand, the variance $ \Sigma_{\scriptscriptstyle IJ}^2 =  
        \left\langle S_{\scriptscriptstyle IJ}^2 \right\rangle 
     -  \left\langle S_{\scriptscriptstyle IJ}\right\rangle ^2 $ 
is related to the root-mean-square amplitudes of noises which give dominant contributions in the weak-signal limit. 
Thus, Eq.~(\ref{eq:def_SNR}) naturally represents the SNR. 
Note that the factor $(T_{\rm obs}/\tau)^{1/2}$ arises from the assumption that the output signal $S_{\scriptscriptstyle IJ}$ in each chunk data can be treated as a statistically independent variable. 
Furthermore, the amplitude of $\mathrm{SNR}$ depends on the optimal filter function. In order to increase the sensitivity to the gravitational-wave signals, an appropriate functional form of the filter function $Q(t)$ or its Fourier counter-part $\tilde{Q}(f)$ must be specified, which we will focus on later.

The definition (\ref{eq:def_SNR}) provides a useful measure to discuss the detectability of isotropic GWBs. On the other hand, when the sky distribution of GWB is anisotropic, it is more convenient to introduce an alternative measure to explore the detection of anisotropic components in GWBs. 
To do this, first notice that the signal 
$S_{\scriptscriptstyle IJ}$ cannot be rigorously treated as stationary random variables. 
As the detector sweeps across the sky, the observed amplitude of GWBs varies in time, since the detector's sensitivity to GWs has a strong directional dependence (See \cite{Taruya:2005yf} for an example of time variation of output signal). 
For instance, when the detector orbits around the Sun with the period $T_{\rm orbit}$, the ensemble average 
$\langle S_{\scriptscriptstyle IJ} \rangle$ also has periodicity in time, 
and one can expand it with frequency $\omega=2\pi/T_{\rm orbit}$ as 
\begin{eqnarray}
  \langle S_{\scriptscriptstyle IJ}(t)\rangle = 
\sum_{m=-\infty}^{+\infty}\,\,\langle S_{{\scriptscriptstyle IJ},m} 
\rangle \,\,e^{i\,m \,\omega\,t}.
\end{eqnarray} 
Thus, the information about anisotropies is encoded in the coefficient 
$\langle S_{{\scriptscriptstyle IJ},m} \rangle$ ~($m\neq0$), and it can be detected if $m\neq0$ component is sufficiently large compared to the noise contributions. 
Accordingly, in similar manner to the isotropic case, one can define the SNR by 
\begin{eqnarray}
\left(\mathrm{SNR} \right)_m = 
\left(\frac{T_{\rm obs}}{T_{\rm orbit}}\right)^{1/2}\,\,
\frac{\mu_{{\scriptscriptstyle IJ},m}}{\Sigma_{{\scriptscriptstyle IJ},m} }
    \equiv \left(\frac{T_{\rm obs}}{T_{\rm orbit}}\right)^{1/2}\,\,
\frac{\left|\langle S_{{\scriptscriptstyle IJ},m}\rangle\right| }{  
    \sqrt{
\big\langle \left|S_{{\scriptscriptstyle IJ},m}\right|^2  \big\rangle 
     - \left|\left\langle S_{{\scriptscriptstyle IJ},m}\right\rangle\right|^2}
     }.
\label{eq:def_SNR2}
\end{eqnarray}
Here, the factor $(T_{\rm obs}/T_{\rm orbit})^{1/2}$ differs from the one in the isotropic case (\ref{eq:def_SNR}), since the signal $S_{{\scriptscriptstyle IJ},m}$ should be evaluated in each orbital period and statistically independence is only achieved between the variables measured at different period.

\section{Optimal filter function and 
signal-to-noise ratio}
\label{sec:optimal_filter}

\subsection{Isotropic signals}
\label{subsec:isotropic}

In this section, we discuss how to choose an optimal filter function to increase the SNRs of GWB in both isotropic and anisotropic cases. Consider first the isotropic GWBs in which the power spectral density $S_h$ is characterized only by the frequency $f$. 
In this case, the task is to calculate the quantities 
$\mu_{\scriptscriptstyle IJ}$ and $\Sigma_{\scriptscriptstyle IJ}$ and determine $\tilde{Q}(f)$ so as to maximize the SNR. 
In appendix \ref{appendix:derivation}, owing to the Gaussian assumption, the statistical quantities are calculated and the results are presented there. Here, we only quote the final expressions.

Under the assumption that the two different detectors (or output data stream) have no correlation of noise, the mean value of the output $\mu_{\scriptscriptstyle IJ}$ becomes  
\begin{equation}
\mu_{\scriptscriptstyle IJ}= \tau \,\int_{-\infty}^{\infty} 
\frac{df}{2} 
\widetilde{Q}(f)
\Bigl[
  C_{\scriptscriptstyle IJ} (f;t,t)+
  \delta_{\scriptscriptstyle IJ}\,N_{\scriptscriptstyle I}(f)
\Bigr],
\label{eq:mu}
\end{equation}
where the quantity $C_{\scriptscriptstyle IJ} (f;t,t)$ is defined as 
\begin{equation}
C_{\scriptscriptstyle IJ} (f;t,t)= \int \frac{d\mathbf{\Omega}}{4\pi}\,
S_h(f)\, \mathcal{F}_{\scriptscriptstyle IJ}(f,\mathbf{\Omega};t, t). 
\label{eq:corr_C} 
\end{equation}
The quantity $\mathcal{F}_{\scriptscriptstyle IJ}$ denotes the antenna pattern function (see Eq.(\ref{eq:def_antenna_pattern})). 
In the case of isotropic GWBs, the so-called overlap-reduction function $\gamma(f)$ can be defined and is related with 
$\mathcal{F}_{\scriptscriptstyle IJ}$ 
(e.g., \cite{Flanagan:1993ix,Allen:1997ad}):  
\begin{equation}
\gamma_{\scriptscriptstyle IJ}(f) 
=\frac{5}{2}\,\,\int\frac{d\mathbf{\Omega}}{4\pi}\,
\mathcal{F}_{\scriptscriptstyle IJ}(f,\mathbf{\Omega};t,t).
\label{eq:overlap_reduc_func}
\end{equation}
The explicit functional form of the antenna pattern function is determined by the detector tensor $D_{\scriptscriptstyle I}^{ij}$ as well as the motion of gravitational-wave detectors (see Appendix 
\ref{appendix:antenna}). Note that while the time-dependence appears in the above expression, the final output $C_{\scriptscriptstyle IJ}$ itself is statistically independent of time in the case of isotropic GWBs. Hence, we omit the time dependence and simply denote 
$C_{\scriptscriptstyle IJ}(f)$ hereafter.

The general expression for $\Sigma_{\scriptscriptstyle IJ}$ is rather complicated and requires a lengthy calculation. 
Hence, we first focus on the {\it weak-signal limit} as the simplest case, in which the amplitude of instrumental noises is assumed to be large compared to that of the gravitational-wave signals, i.e., 
$h_{\scriptscriptstyle I}(t)\ll n_{\scriptscriptstyle I}(t)$.  
In this case, the squared quantity $\Sigma^2_{\scriptscriptstyle IJ}$ is reduced to 
\begin{equation}
\Sigma^2_{\scriptscriptstyle IJ}\simeq 
\langle S_{\scriptscriptstyle IJ}^2\rangle \simeq
\frac{\tau}{2}\,
\int_{-\infty}^{\infty} \frac{df}{2}\,\,
\left|\widetilde{Q}(f)\right|^2 \,\,
N_{\scriptscriptstyle I}(f)N_{\scriptscriptstyle J}(f). 
\label{eq:sigma2_weak}
\end{equation}
Based on the expressions (\ref{eq:mu}) and (\ref{eq:sigma2_weak}), let us determine the functional form of $\widetilde{Q}(f)$ that maximizes 
$\mathrm{SNR}$. According to \cite{Allen:1997ad}, this becomes remarkably simple if we introduce an inner product $(A|B)$ for any pair of complex functions $A(f)$ and $B(f)$:
\begin{equation}
(A|B)\equiv \int_{-\infty}^{\infty}\frac{df}{2}\,A^*(f) \, B(f)\,
\,N_{\scriptscriptstyle I}(f)N_{\scriptscriptstyle J}(f).
\label{eq:inner_product1}
\end{equation}
In terms of this, the SNR 
(\ref{eq:def_SNR}) is rewritten in the form: 
\begin{equation}
\mathrm{SNR}^2 
\simeq  \frac{ 2\,\,T_{\rm obs}}
{{\displaystyle (\widetilde{Q}|\widetilde{Q})}  }
\left(
    \widetilde{Q}^*\,\left| 
    \frac{C_{\scriptscriptstyle IJ}(f)}{N_{\scriptscriptstyle I}(f)
    N_{\scriptscriptstyle J}(f)}\right. 
\right)^2 
\quad(I\neq J).
\label{eq:SNR^2_weak}
\end{equation}
Here, we only consider the case $I\neq J$. For the self-correlation signal ($I=J$), the mean value $\mu_{\scriptscriptstyle IJ}$ contains the instrumental noise spectrum and the signal cannot be optimized (see below). 
\footnote{
Furthermore, it is worthwhile to note that for the cross-correlation between so-called $A,E,T$ variables the signal-to-noise ratio vanishes because of $C_{IJ}=0$ \, ($I \neq J,~I=A,E,T$). 
As proved in \cite{Kudoh:2004he}, this property is the general consequences of the spacecraft configuration, and then it holds true irrespective of the choice of filter function. 
This is a kind of null signal stream for a stochastic background sources. 
}
Since $(A|B)$ satisfies the same properties as held for an ordinary inner product of vectors in three-dimensional Euclidean space, the norm is positive-definite. Then the problem to determine the filer function is analogous to find the vector $\vec{Q}$ that maximize the quantity $(\vec{Q}\cdot\vec{A})^2/(\vec{Q}\cdot\vec{Q})$. 
The answer is 
\begin{equation}
\tilde{Q}(f)= c\,\,\frac{C_{\scriptscriptstyle IJ}^*(f)}
{N_{\scriptscriptstyle I}(f)N_{\scriptscriptstyle J}(f)}, 
\label{eq:filer1}
\end{equation}
where $c$ is merely an arbitrary constant. 
The resultant filter function $\widetilde{Q}(f)$ depends on the spectrum of GWB as well as the antenna pattern function and the instrumental noises. Thus, within the bandwidth for interest of gravitational-wave detector, some templates of the spectrum $S_h(f)$ is needed to detect the GWBs. 
Substituting (\ref{eq:filer1}) into (\ref{eq:SNR^2_weak}), the SNR is finally obtained in the form \cite{Allen:1997ad}:  
\begin{equation}  
\mathrm{SNR}\simeq \sqrt{2\,\,T_{\rm obs}}\,\,
\left[\int_{-\infty}^{\infty} \frac{df}{2}\,\,
\frac{|C_{\scriptscriptstyle IJ}(f)|^2}
{N_{\scriptscriptstyle I}(f)N_{\scriptscriptstyle J}(f)}\right]^{1/2} 
\quad (I\neq J, \quad h_I \ll n_I ).
\label{eq:SNR_weak}
\end{equation}

The above results are now well known in the literature and have been currently used in the data analysis of GWB searches, because most of the candidate for GWBs is far below the noise sensitivity of ground-based detectors. 
In general situations with large amplitudes of GWBs, which may be the case for the next-generation space interferometers, there appear the additional terms that contribute to the variance 
$\Sigma_{\scriptscriptstyle IJ}^2$. Their final expression is 
\footnote{
By construction, the variance $\Sigma^2_{IJ}$ should be a positive definite function. 
This consistency is explicitly satisfied for self-correlation because $N_I(f)$ and $C_{II}$ are positive definite real functions. 
However, for cross-correlation, positive definiteness of the variance seems non-trivial due to the first term in the integrand (\ref{eq:sigma2_general}), i.e., $(C_{IJ})^2$,  which is in general a complex variable. 
This term comes out after the application of the Wick theorem (Appendix \ref{appendix:derivation}).   
(Note that this is not a problem for ground-based detectors because the antenna pattern functions of ground-based detectors are real functions.) 
The positive definiteness can be shown as follows. 
If the inequality
$$
{\mathcal{I}} = \int^\infty_{-\infty} \frac{df}{2} 
\biggl\{
 \widetilde{Q}^2  C_{IJ} ^2
 +
 |\widetilde{Q}|^2 
       C_{II}  C_{JJ} 
\biggr\}  \ge 0 
\label{eq:positive def1}
$$
is proved to be satisfied, then the positive definiteness is explicit. 
To prove this inequality, we consider the complex function $C_{IJ}$ as an inner product for complex vectors defined by ${\mathbf{F}}_I = e^{-i 2\pi f {\mathbf{\Omega  \cdot  x}_I}} (F_I^+, F_I^\times)$.
Then, one finds an inequality
$
 {\mathcal  I} 
\ge
      2 \int^\infty_{0} df ~ |\widetilde{Q}|^2  \left[
     C_{II} C_{JJ} - |C_{IJ}|^2
\right], 
$
where we have used the fact that $ -2 |z|^2 \le z^2+z_*^2 \le 2 |z|^2$ and $|z_1 z_2|=|z_1||z_2|$ for complex variables. 
By applying the Schwarz's inequality $|C_{IJ}|^2 \le  {C_{II}} {C_{JJ}}$, the above inequality is shown to be satisfied, and the proof is completed. 
}
\begin{equation}
\Sigma^2_{\scriptscriptstyle IJ}= \frac{\tau}{2}\,
\int_{-\infty}^{\infty} \frac{df}{2}
\left[
\widetilde{Q}^2(f) \,\,V(f) + \left|\widetilde{Q}(f)\right|^2 \,\,W(f) 
\right],
\label{eq:sigma2_general}
\end{equation}
where the functions $V(f)$ and $W(f)$ are defined by:  
\begin{eqnarray}
U(f)&=& C_{\scriptscriptstyle IJ}(f) +  
\delta_{\scriptscriptstyle IJ} \,N_{\scriptscriptstyle I}(f),
\nonumber
\\
V(f)&=& U^2= 
\bigl[ C_{\scriptscriptstyle IJ} \bigr]^2 
+ 
 \delta_{\scriptscriptstyle IJ} 
 N_{\scriptscriptstyle I} 
 \Bigl[  N_{\scriptscriptstyle I} 
        +  2C_{\scriptscriptstyle II} 
 \Bigr],
\nonumber
\\
W(f)&=&
C_{\scriptscriptstyle II} C_{\scriptscriptstyle JJ} + 
C_{\scriptscriptstyle II} N_{\scriptscriptstyle J}  + 
C_{\scriptscriptstyle JJ} N_{\scriptscriptstyle I}  + 
N_{\scriptscriptstyle I} N_{\scriptscriptstyle J} .   
\nonumber
\end{eqnarray}
In the above expressions, $W$ is a real function, while $U$ and $V$ are generally complex functions of $f$. 
Note that this situation is somewhat different from those considered by Allen \& Romano \cite{Allen:1997ad}. 
In their paper, the quantity $C_{\scriptscriptstyle IJ}$ is assumed to be real, since they particularly focused on the ground-based detectors of Fabry-Perot type. 
Then the overlap-reduction function $\gamma(f)$ is a real function of 
$f$ and accordingly the functions $U$ and $V $ becomes real. 
The integrand in right-hand-side of Eq.~(\ref{eq:sigma2_general}) may be factorized as $|Q|^2\{V+W\}$ and the explicit functional form of the optimal filter is derived just following the same procedure as discussed in the weak-signal limit.

For space-based detectors whose signal extraction method relies upon the Doppler tracking technique, 
$C_{\scriptscriptstyle IJ}(f)$ is not necessarily a real function because the overlap-reduction function $\gamma(f)$ sometimes becomes complex.  
This generally happens unless the relation 
$\mathcal{F}_{\scriptscriptstyle IJ}(f,\mathbf{\Omega})=
\mathcal{F}^*_{\scriptscriptstyle IJ}(f,-\mathbf{\Omega})$ holds for the antenna pattern function, depending on both the detector tensor 
$D_{\scriptscriptstyle I}^{ij}$ and the specific combination of time-delayed signals
\footnote{
For cross-correlation analysis ($I \neq J$), the relation 
$\mathcal{F}_{\scriptscriptstyle IJ}(f,\mathbf{\Omega})=
\mathcal{F}^*_{\scriptscriptstyle IJ}(f,-\mathbf{\Omega})$ holds 
only among the same types of (first-generation) TDI variables. 
If we consider a cross-correlation between different type of TDI variables, for example between Sagnac variables and so-called $X$ variables, the relation is not true. (This type of cross-correlation is possible for a hexagonal space-interferometer without introducing correlated noises.)
Furthermore, for much complicated time-dependent TDI variables like the second-generation \cite{Shaddock:2003dj,Tinto:2003vj}, the relation does not hold anymore in general.
}.
In that case, the quantity $\mathrm{SNR}$ cannot be simply expressed by using the inner product (\ref{eq:inner_product1}). 
Here, to generalize the above-mentioned procedure, instead of using 
(\ref{eq:inner_product1}), we introduce a new inner product: 
\begin{equation}
    \left\{A|B\right\}\equiv \int_{0}^{\infty}\frac{df}{2}\,\,
    \mathbf{A}^{\dagger}(f)\cdot \mathbf{M}(f)\cdot \mathbf{B}(f),
    \label{eq:inner_product2}
\end{equation}
where $\mathbf{A}$ and $\mathbf{B}$ represent two-dimensional vectors, whose component is defined by $\mathbf{A}=(A,A^*)$. 
The complex matrix  $\mathbf{M}$ is given by 
\begin{equation}
    \mathbf{M}(f)=
    \left(
\begin{array}{cc}
W(f) & V^*(f) \\
V(f) & W(f) 
\end{array}
\right).
\end{equation}
Recalling the fact that $V^*(f)=V(-f)$ and 
$\widetilde{Q}^*(f)=\widetilde{Q}(-f)$, Eq.~(\ref{eq:sigma2_general}) is rewritten as 
$\Sigma_{\scriptscriptstyle IJ}^2=(\tau/2)\,
\{\widetilde{Q}|\widetilde{Q}\}$. 
Similarly, the mean value 
$\mu_{\scriptscriptstyle IJ}$ becomes
\begin{eqnarray}
    \mu_{\scriptscriptstyle IJ}
    =
    \tau \int_0^{\infty}\frac{df}{2}
    \left[ U(f)\widetilde{Q}(f) + U^*(f)\widetilde{Q}^*(f) \right]
    = \tau\,\,
    \left\{ \frac{U\,V^*-U^*W}{|V|^2-W} \biggr| ~ \widetilde{Q}\right\}.
    \nonumber
\end{eqnarray}
The squared quantity $\mathrm{SNR}$ is thus expressed in a closed form using the new inner product: 
\begin{eqnarray}
\mathrm{SNR}^2 
= 
\frac{2\,\,T_{\rm obs}}
{ \bigl\{ \widetilde{Q} \, \bigr|  \widetilde{Q}  \bigr\}  } 
\left\{
     \frac{U V^*-U^* W}{|V|^2-W^2} \biggr| \widetilde{Q}
\right\}^2.  
\label{eq:SNR general,isotropic,Q}
\end{eqnarray}
Since $\{A|B\}$ satisfies the same properties as held for $(A|B)$, it is now straightforward to determine the optimal filter: 
\begin{equation}
    \tilde{Q}(f)= c \,\,\frac{U(f) V^*(f)-U^*(f) W(f)}{|V(f)|^2-W^2(f)}. 
    \label{eq:filter2}
\end{equation}
The filter function (\ref{eq:filter2}) is a generalization of the result in weak-signal case. This is also an extension of the result given by 
Allen \& Romano \cite{Allen:1997ad} (c.f. Eq.~(5.13) of their paper) to the situations in which  $C_{\scriptscriptstyle IJ}(f)$ is complex. 
With the new optimal filter, Eq.~(\ref{eq:SNR general,isotropic,Q}) leads to
\begin{equation}
  \mathrm{SNR}=\,\,\sqrt{2 T_{\mathrm{obs}}} \,
  \left[\int_{-\infty}^{\infty}\frac{df}{2}\,\, \frac{|U(f)|^2}{|U(f)|^2+W(f)} 
  \right]^{1/2}.
\label{eq:SNR_general}
\end{equation}
It is worthwhile to note that the SNR 
$\mathrm{SNR}$ as well as the optimal filter $\widetilde{Q}(f)$ gives a meaningful definition only in the cross-correlation case 
($I\neq J$). For the self-correlation $(I=J)$, the function 
$U(f)$ becomes real and the relation $V=W=U^2$ holds, which makes the expression (\ref{eq:filter2}) ill-defined. This means that the SNR for self-correlation signal cannot be maximized by the filter function and it should be simply given by the ratio $C_{\scriptscriptstyle II}(f)/N_{\scriptscriptstyle I}(f)$. 
Accordingly, the dependence of the observational time is dropped and SNR does not increase in time. 
This is a natural consequence and is even true in the weak-signal limit [see Eq. (\ref{eq:SNR_weak})].

Finally, we note that the results (\ref{eq:filter2}) and (\ref{eq:SNR_general}) can be rewritten in terms of the overlap reduction function (\ref{eq:overlap_reduc_func}) as 
\begin{eqnarray}
Q(f) &=& \frac{2}{5} \,\frac{S_h(f)\gamma^*_{\scriptscriptstyle IJ}(f)}{R(f)},
\label{eq:filter_using_gamma}
\\
\mathrm{SNR}
  &=& \sqrt{2 T_{\mathrm{obs}} }
\left[\int_{-\infty}^{\infty}\frac{df}{2}
\,
\left(\frac{2}{5}\right)^2 
\frac{S_h^2(f)\left|\gamma_{\scriptscriptstyle IJ}(f)\right|^2}
{R(f)}\right]^{1/2}, 
\label{eq:SNR_using_gamma}
\end{eqnarray}
with the function $R(f)$ being 
\begin{eqnarray}
R(f) 
= \left(\frac{2}{5}\right)^2 S_h^2(f) \,\,
  \Bigl( \left|\gamma_{\scriptscriptstyle IJ} \right|^2
   +
  \gamma_{\scriptscriptstyle  II}\, \gamma_{\scriptscriptstyle JJ}
\Bigr)
+
  \frac{2}{5} S_h(f) 
  \Bigl( \gamma_{\scriptscriptstyle II}\, N_{\scriptscriptstyle J}
   + \gamma_{\scriptscriptstyle JJ}(f)\, N_{\scriptscriptstyle I} 
 \Bigr)
 + N_{\scriptscriptstyle I}\, N_{\scriptscriptstyle J}.
\end{eqnarray}
Compared with the results in Ref. \cite{Allen:1997ad} (see Sec.~5-A of their paper), the above expressions are more general and are also applicable to the cases with ${\mathrm{Im}} [\gamma_{\scriptscriptstyle IJ}]\ne0$ or 
$\gamma_{\scriptscriptstyle II}\ne0$. 
%
%
%
%
%
%
\subsection{Anisotropic signals}
\label{subsec:anisotropic}
%
%
%
%
%
%
%
%
We next discuss the anisotropies of GWB, which may be a key ingredient to discriminate between the cosmological origin and the Galactic origin of GWBs. In the presence of anisotropies, the $m\ne0$ components of the coefficients $\mu_{{\scriptscriptstyle IJ},m}$ and 
$\Sigma_{{\scriptscriptstyle IJ},m}$ given in (\ref{eq:def_SNR2}) become non-vanishing. The important point to emphasize is that the actual values of  $m\ne0$ components is not only determined by the 
GWB signal $S_h(f,\mathbf{\Omega})$, but also by the angular response of gravitational-wave detectors as well as the instrumental noises.

While the expression of $\mu_{{\scriptscriptstyle IJ},m}$ is easy to derive, a full expression of $\Sigma_{{\scriptscriptstyle IJ},m}$ requires a rather lengthy calculation, together with some approximations. 
Details of the calculation are described in Appendix 
\ref{appendix:derivation}. The final results are 
\begin{eqnarray}
\mu_{{\scriptscriptstyle IJ},m} 
&=& 
\tau\,\left|\int_{-\infty}^{\infty} 
\frac{df}{2}\,\,
\widetilde{Q}(f) 
\Bigl[ 
    C_{{\scriptscriptstyle IJ},m} (f) +
    \delta_{m0}\delta_{\scriptscriptstyle IJ}\,N_{\scriptscriptstyle I}(f)
\Bigr] 
\right|,  
\label{eq:mu_m}
\\
\Sigma_{{\scriptscriptstyle IJ},m}^2
&=&
\frac{\tau}{2}\,\left(\frac{T_*}{T_{\rm orbit}}\right)\,
\int_{-\infty}^{\infty}\frac{df}{2}
\left[
\widetilde{Q}^2(f) \,\,V_m(f) + \left|\widetilde{Q}(f)\right|^2 \,\,W_m(f) 
\right], 
\label{eq:sigma2_mu_general}
\end{eqnarray}
where the functions $V_m(f)$ and $W_m(f)$ are given by 
\begin{eqnarray}
V_m(f)&=& \sum_n C_{{\scriptscriptstyle IJ},n} 
C_{{\scriptscriptstyle IJ},-n}   +\delta{\scriptscriptstyle IJ}
\left[ 2 C_{{\scriptscriptstyle II},0} 
         N_{\scriptscriptstyle I} + 
\left(\frac{\tau}{T_*}\right)^2 N_{\scriptscriptstyle I}^2 \right],
\nonumber
\\
W_m(f)&=&  \sum_n C_{{\scriptscriptstyle II},n} 
C_{{\scriptscriptstyle JJ},-n}  + 
C_{{\scriptscriptstyle II},0} N_{\scriptscriptstyle J}  + 
C_{{\scriptscriptstyle JJ},0} N_{\scriptscriptstyle I}  + 
\left(\frac{\tau}{T_*}\right)\,
N_{\scriptscriptstyle I} N_{\scriptscriptstyle J}  , 
\nonumber
\end{eqnarray}
with the correlation signal $C_{{\scriptscriptstyle IJ},m}(f)$ being 
\begin{equation}
    C_{{\scriptscriptstyle IJ},m}(f) = \frac{1}{T_{\rm orbit}}\,\,
    \int_0^{T_{\rm orbit}}dt\,\,e^{-i\,m\,\omega\,t}\,
    C_{\scriptscriptstyle IJ}(f;t,t).
    \label{eq:def_of_CIJ_m}
\end{equation}
Here, the timescale $T_*$ has been introduced to characterize the signal correlations observed at the different times. 
Roughly speaking, it is inversely proportional to the velocity of detector (spacecraft) times the observed frequency, i.e., 
$T_*=(2\pi \dot{x}_{\scriptscriptstyle I}\,f)^{-1}$.
The above expression is valid when the local observation time $\tau$ is sufficiently longer than $T_*$.

The full expression (\ref{eq:sigma2_mu_general}) implies that the $m=0$ component of the GWB signal as well as the other contribution of anisotropic signals may act as a disturbance, which reduces the SNR of anisotropic GWB. 
This might be crucial for the detection of anisotropies in the strong-signal case, which is indeed the case considered in \cite{Taruya:2005yf}. 
On the other hand, for the next-generation space interferometers, the observational frequency band is around 
$f\sim0.1-10$ Hz and the dominant sources of GWB are extragalactic and cosmological origin. Thus, the anisotropic components are expected to be very weak. While several GW sources could produce large amplitude of (isotropic) GWB, the anisotropies of them are still small. 
For this reason, we hereafter focus on the weak-signal cases, and write down the expressions for the SNR as well as the optimal filter in the cases.

Taking the weak-signal limit, Eq.~(\ref{eq:sigma2_mu_general}) now reduces to
\begin{equation}
\Sigma_{{\scriptscriptstyle IJ},m}^2 \simeq
\left\langle \left|S_{{\scriptscriptstyle IJ},m}\right|^2\right\rangle
\simeq \frac{\tau^2}{2\, T_{\rm orbit}}\,
\int_{-\infty}^{\infty}\frac{df}{2}
\,\, \left|\widetilde{Q}(f)\right|^2 \,\,
N_{\scriptscriptstyle I}(f)N_{\scriptscriptstyle J}(f).
\label{eq:sigma2_mu_weak}
\end{equation}
This expression is very similar to Eq.~(\ref{eq:sigma2_weak}) in the isotropic case. Thus, one may use the same definition of inner product as defined in Eq.~(\ref{eq:inner_product1}) to express the SNR (\ref{eq:def_SNR2}). We then have
\begin{eqnarray}
\left(\mathrm{SNR} \right)_m 
\simeq
   \sqrt{ \frac{ 2 T_{\rm obs} }{ (\widetilde{Q}|\widetilde{Q})} }
\left| \left( 
        \widetilde{Q}^*\,\left|\,
        \frac{C_{{\scriptscriptstyle IJ},m}(f)}{N_{\scriptscriptstyle I}(f)
        N_{\scriptscriptstyle J}(f)}\right.
\right)\right| 
\quad(I\neq J).
\end{eqnarray}
The optimal filter for $m$-th component is 
$
\widetilde{Q}_m \propto  C_{{\scriptscriptstyle IJ},m} /
{N_{\scriptscriptstyle I} N_{\scriptscriptstyle J} }
$, 
and the resultant expression for SNR is 
\begin{equation}
\left(\mathrm{SNR} \right)_m \simeq
 \sqrt{2 T_{\rm obs}} 
\left[\int_{-\infty}^{\infty} \frac{df}{2}\,\,
\frac{|C_{{\scriptscriptstyle IJ},m}(f)|^2}
{N_{\scriptscriptstyle I}(f)N_{\scriptscriptstyle J}(f)}\right]^{1/2} 
\quad (I\neq J , ~ h_I \ll n_I). 
\label{eq:SNR2_weak}
\end{equation}
This expression is the same equation as first derived by 
Allen \& Ottewill \cite{Allen:1997gp} and has been frequently used in the literature. In the next section, we will use Eq.~(\ref{eq:SNR2_weak}) to discuss the detectability of dipole anisotropy induced by the proper motion of our local system.

\section{Sensitivity of next-generation space interferometers to stochastic GWBs}
\label{sec:sensitivity}

\subsection{Interferometer design}
\label{subsec:design}

Currently, practical interferometer design as well as precise orbital configurations for proposed future missions are still under debate and are not fixed. 
Nevertheless, it is commonly accepted that the next-generation space interferometers will aim at detecting the GWB of primordial origin generated during the inflationary epoch. 
In this respect, the low-frequency band that is not compromised by the astrophysical foregrounds is thought to be favorable and the frequency around $0.1-1$ Hz would be the best observational window. 
Based on these, we consider several types of interferometric design and discuss prospects for the detectability of primordial backgrounds owing to the correlation analysis.

\begin{figure}[tb]
\begin{center}
\includegraphics[width=8cm,clip]{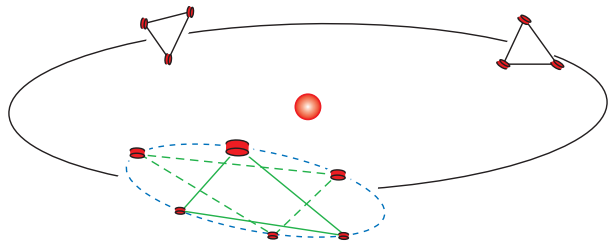}
\caption{
\label{fig:hexagonal interferometer}
A typical orbital configuration for a future space interferometer. 
One of the three interferometers on the ecliptic orbit consists of six spacecrafts, and the six probes form a hexagonal space-interferometer. 
}
\end{center}
\end{figure}

Fig.~\ref{fig:hexagonal interferometer} shows an example of the constellation of the spacecrafts as well as the orbital configuration of space interferometers. 
We assume that the future missions consists of four sets of detectors; two of which consist of three spacecrafts forming a triangular configuration, like LISA, and the remaining of which consists of six spacecrafts forming a star-like constellation. 
Each of the three detectors are located separately $120^{\circ}$ ahead or behind on the orbit around the Sun. With this setup, we now consider the three possible cases summarized in Table \ref{table: instrument parameter}. 
For comparison, we also list the instrumental parameters of LISA.

The original DECIGO (or BBO) is planning to use Doppler-tracking method as will be implemented in LISA. 
In such signal processing technique, the output signals sensitive to gravitational waves are constructed by time-delayed combination of laser pulses to cancel out laser frequency noises.  
This type of interferometry is known as time-delay interferometry (TDI). 
    To distinguish the original DECIGO from FP-DECIGO, we call the original one TDI-DECIGO. 
To examine the detector sensitivity to both the isotropic and anisotropic 
GWBs, we use the Michelson-like TDI variable called $X$ ($Y,Z$) variables \cite{Armstrong:1999} (see Appendix \ref{appendix:antenna}). 
With this specific data stream, we perform the cross-correlation analysis using the signals extracted from the spacecrafts forming a star-like configuration. 
On the other hand, the signal processing of the space interferometer 
FP-DECIGO may adopt the same technique as used in the ground detectors.
The essential requirement is that the relative displacement between the spacecrafts to be constant during an observation. 
Adopting the Fabry-Perot configuration, while the arm-length of the detector can be greatly reduced without changing the observed frequency range, no flexible combination of time-delayed signal is possible anymore. 
We assume that the output data which is available for data analysis is only one for each set of detectors. 
The third class of space interferometer discussed in this paper is the ultimate DECIGO. It is an extreme version of TDI-DECIGO, whose signal sensitivity is limited only by quantum noises. 
Although the significant technological developments are necessary to achieve the ultimate sensitivity, we intend to consider it as an observational limitation.

\begin{figure}[t]
\begin{center}
\includegraphics[width=10cm,clip]{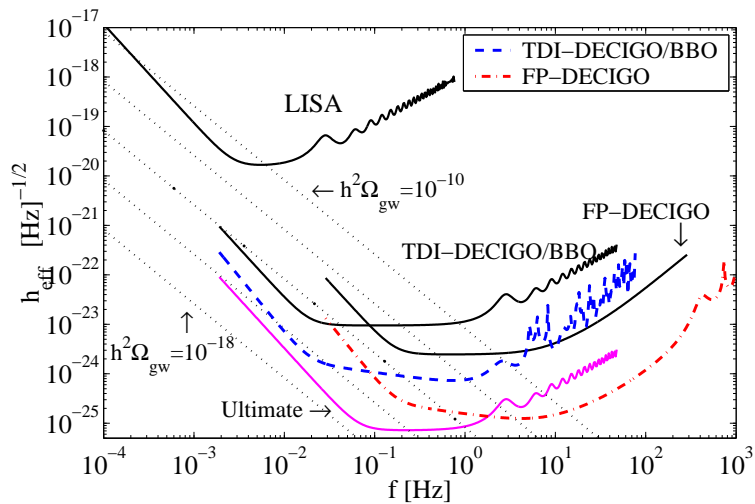}
\caption{
\label{fig:h_eff for omega_GW}
Spectral amplitude sensitivity $h_{\mathrm{eff}}$ for several space-based interferometers~\cite{Larson:1999we}. 
Solid curves show the sensitivity for self-correlation analysis for LISA, 
TDI-DECIGO/BBO, FP-DECIGO and ultimate DECIGO 
(Table \ref{table: instrument parameter}). 
So-called $X$-variable \cite{Armstrong:1999} of TDIs is used for the self-correlation analysis, and the transfer function of 
FP-DECIGO can be found in, e.g., \cite{Schilling:1997id}. 
The dashed curves show the sensitivity for cross-correlation analysis, assuming the hexagonal spacecraft configuration.
Oscillation of the sensitivity curves at high frequency band comes from overlap reduction function. 
Dotted lines show $h^2 \ogw = 10^{-10}, 10^{-12}, \cdots, 10^{-18}$. 
In these plots, we have taken $\overline{\mathrm{SNR}}=5$, $\Delta f=f/10$, and $T_{\rm obs}=1$ year.  
}
\end{center}
\end{figure}

In Fig.~\ref{fig:h_eff for omega_GW}, the solid lines show the sensitivity curves for four types of interferometers. 
Compared to TDI-DECIGO/BBO and ultimate DECIGO, the best sensitivity of FP-DECIGO is slightly shifted to the higher frequency band, $f \sim 0.5-10$Hz. 
As a result, while the strain amplitude of FP-DECIGO has even better sensitivity than that of TDI-DECIGO/BBO, it conversely becomes worse when quantifying the sensitivity by means of $\ogw$. Note, however, that the sensitivity curves plotted here just represent the noise intensity in comparison with the detector response and do not correctly reflect the detection limit of GWB. The quantitative aspect of the detectability of GWB should be investigated through the cross-correlation analysis, which we will discuss in detail.

\begin{table}[tb]
\begin{ruledtabular}
\begin{tabular}{ccccl}
 &$L [{\rm m}]$ & $S_{\rm shot}[{\rm m }\,{\rm Hz}^{-1/2}]$ & 
   $S_{\rm accel} [{\rm m}\,{\rm s}^{-2}\,{\rm Hz}^{-1/2}]$&  
   interferometric type 
\\ \hline  \hline  
LISA & $5\times 10^9$ & $2\times 10^{-11}$ & $3\times 10^{-15}$ & 
Doppler tracking
\\
TDI-DECIGO/BBO& $5\times 10^7$ & $1.2\times 10^{-16}$ & $3.9\times 10^{-17}$ & 
Doppler-tracking
\\
FP-DECIGO & $1\times 10^6$ & $2.2\times 10^{-18}$ & $7.9\times 10^{-19}$ & 
Fabry-Perot with fineness ${\mathcal F}=10$ 
\\
Ultimate DECIGO & $5\times 10^7$ & $3\times 10^{-19}$ & $3 \times 10^{-19}$ & 
Doppler-tracking limited by quantum noise
\\
\end{tabular}
\end{ruledtabular}
\caption[short]{
Instrumental parameters for next-generation space interferometers (see also Appendix \ref{appendix:antenna}). 
Radiation pressure noise of FP-DECIGO is set to 
$S_{rad}=6 \times 10^{-26} \, f^{-2} [{1+(f/f_0)^2}]^{-1/2} ~ 
{\mathrm Hz}^{-1/2} $ where $f_0$ is given by $f_0 = c/4{\mathcal{F}}L$, and the fineness is ${\mathcal{F}}=10$. 
The shot-noise should also accompany the cutoff frequency, like 
$S_{\mathrm{shot}} \propto [{1+(f/f_0)^2}]^{-1/2} $.
Ultimate DECIGO is an ultimate GW observatory whose sensitivity is only limited by the standard quantum limit, and its effective noise level used in this paper is listed in the table.
For a $100$ Kg mass and an arm length of that of TDI-DECIGO/BBO, the spectral amplitude of the noise could be 
$\sim 10^{-26} \mathrm{Hz}^{-1/2}$ around 0.1 Hz. 
The sensitivity curves are shown in Fig. \ref{fig:h_eff for omega_GW}. 
Note that the instrumental parameters in the table do not necessarily represent latest mission plans.
The mission design of BBO adopts different instrumental parameters. 
}
\label{table: instrument parameter}
\end{table}

\subsection{Isotropic case}
\label{subsec:isotropic_GWB}

To see how the cross-correlation analysis improves the sensitivity to the GWB, let us first focus on the isotropic GWB and evaluate the sensitivity of each space interferometer. 
For this purpose, we take the weak-signal approximation.
From Eq.~(\ref{eq:SNR_weak}) (or Eq.~(\ref{eq:SNR_using_gamma})), we introduce effective strain sensitivity 
\cite{Cornish:2001bb,Kudoh:2004he}:
\begin{equation}
    h_{\rm eff}(f) = \overline{\mathrm{SNR}}^{1/2}\,
    \left\{\frac{5}{2}
    \frac{N_{\scriptscriptstyle I}(f)N_{\scriptscriptstyle J}(f)}
    {|\gamma_{\scriptscriptstyle IJ}(f)|\,
      \Delta f\,\,T_{\rm obs}}\right\}^{1/4}, 
    \label{eq:def_h_eff}
\end{equation}
where the quantity $\overline{\mathrm{SNR}}$ means the signal-to-noise of stochastic GWB 
over the frequency range $f\sim f+\Delta f$. 
Eq. (\ref{eq:def_h_eff}) quantifies the strain amplitude of minimum detectable GWB for different frequency bin. 
Unlike the usual sense of the sensitivity curves, it depends on the observation time as well as the frequency interval.

In Fig.~\ref{fig:h_eff for omega_GW}, the effective sensitivity of cross-correlated signals for TDI-DECIGO/BBO and FP-DECIGO are shown 
({\it dashed} and {\it dot-dashed}). 
Compared to the curves for self-correlation signals, the sensitivity of the cross-correlated signals is greatly improved more than one order of magnitude in the strain amplitude. 
With the signal-to-noise ratio $\overline{\mathrm{SNR}}=5$, the minimum detectable $\ogw$ reaches  
$\ogw\sim 10^{-16}(10^{-15})$ for TDI-DECIGO/BBO (FP-DECIGO).  
Increasing the observational time, the effective sensitivity becomes even comparable to the sensitivity of ultimate DECIGO.

While the effective strain sensitivity $h_{\rm eff}$ provides useful information on the frequency dependence of the detectable amplitude, for more precise estimate of the SNR, one must directly evaluate the expression (\ref{eq:SNR_weak}),  integrating over the whole frequency bins. Assuming the flat spectrum, 
$\ogw(f)=\Omega_{{\rm gw},0}\,f^0$, the resultant detectable values of $\ogw$ are
\begin{equation}
h^2\,\Omega_{{\rm gw},0}=\left\{ 
\begin{array}{cl}
6.8\times10^{-18}
\,\,{\displaystyle \left(\frac{ \mathrm{SNR} }{5}\right)
\left(\frac{T_{\rm obs}}{1 \mathrm{year}}\right)^{-1/2} }
& (\mbox{TDI-DECIGO/BBO}) \\
1.1\times10^{-16}\,\, {\displaystyle
                 \left(\frac{\mathrm{SNR}}{5}\right)
\left(\frac{T_{\rm obs}}{1 \mathrm{year}}\right)^{-1/2} }
& (\mbox{FP-DECIGO}) \\
4.2\times10^{-21}\,\,
{\displaystyle \left(\frac{\mathrm{SNR}}{5}\right)
\left(\frac{T_{\rm obs}}{1 \mathrm{year}}\right)^{-1/2} }
& (\mbox{ultimate DECIGO}) 
\end{array}
\right.
\label{eq:omega_no_cutoff}
\end{equation}
As anticipated from the effective sensitivity curves, the minimum detectable value of $\ogw$ for TDI-DECIGO/BBO is an order of magnitude lower than that for FP-DECIGO. 
In this sense, the frequency band around $0.1-1$Hz covered by 
TDI-DECIGO/BBO may be the best observational window to probe the GWB of primordial origin.

\begin{figure}[tb]
\begin{center}
\includegraphics[width=10cm,clip]{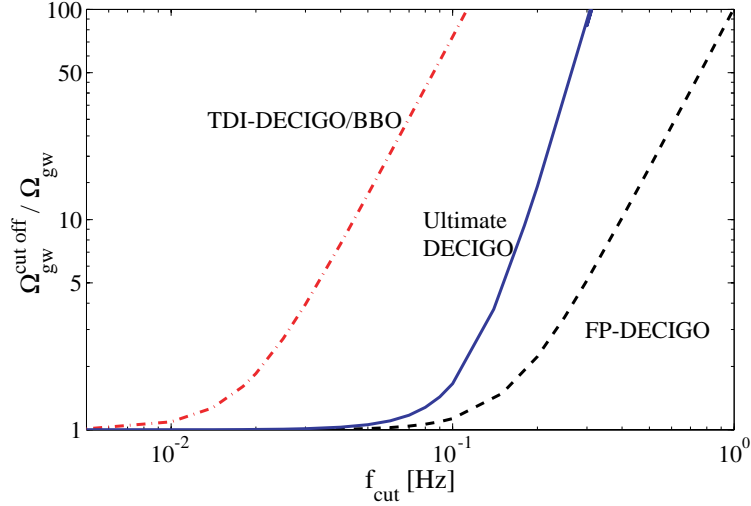}
\caption{
\label{fig:Omega_cutoff} 
Dependence of low-frequency cutoff $f_{\rm cut}$ on the minimum detectable value $\Omega_{\rm gw}$ in the case of flat spectra, $\Omega_{\rm gw}\propto f^0$. 
The resultant values $\Omega_{\rm gw}^{\rm cutoff}$ are plotted by normalizing them to $\Omega_{\rm gw}$ without cutoff.
Observational band (available frequency band) is the same as shown in Fig.~\ref{fig:h_eff for omega_GW}. 
}
\end{center}
\end{figure}

The above discussion is, however, rather optimistic. 
In practice, one would not neglect several astrophysical foregrounds.  
As it has been discussed by several authors 
\cite{Schneider:2000sg,Farmer:2003pa}, a cosmological population of white-dwarf binaries may produce a large signal at low-frequency band $f\lesssim0.2$Hz, which would not be resolved individually. 
Hence, the cosmological white-dwarf binaries may act as a confusion noise and they prevent us from detecting the primordial GWB below the frequency $f_{\rm cut}=0.2$ Hz. 
As recently pointed out by Seto \cite{Seto:2005qy}, the introduction of the low-frequency cutoff in the integral (\ref{eq:SNR_weak}) significantly reduces the SNRs for the GWB signals. 
To see how this affects the detectability quantitatively, we plot the minimum detectable $\ogw$ as a function of the cutoff frequency 
$f_{\rm cut}$ in Fig.~\ref{fig:Omega_cutoff}.

\begin{figure}[tb]
\begin{center}
\includegraphics[width=8.5cm,clip]{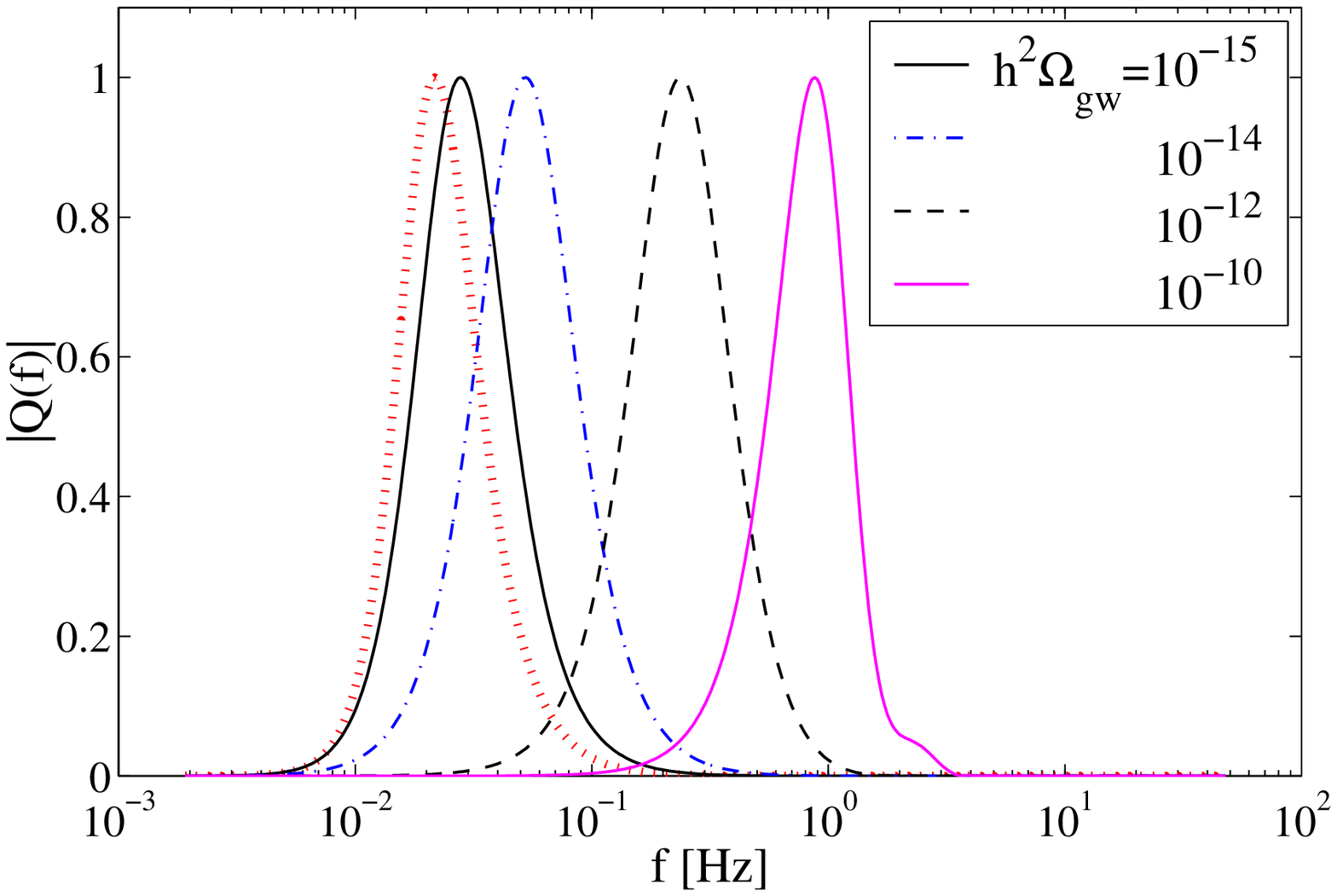}
\hspace*{0.5cm}
\includegraphics[width=8.5cm,clip]{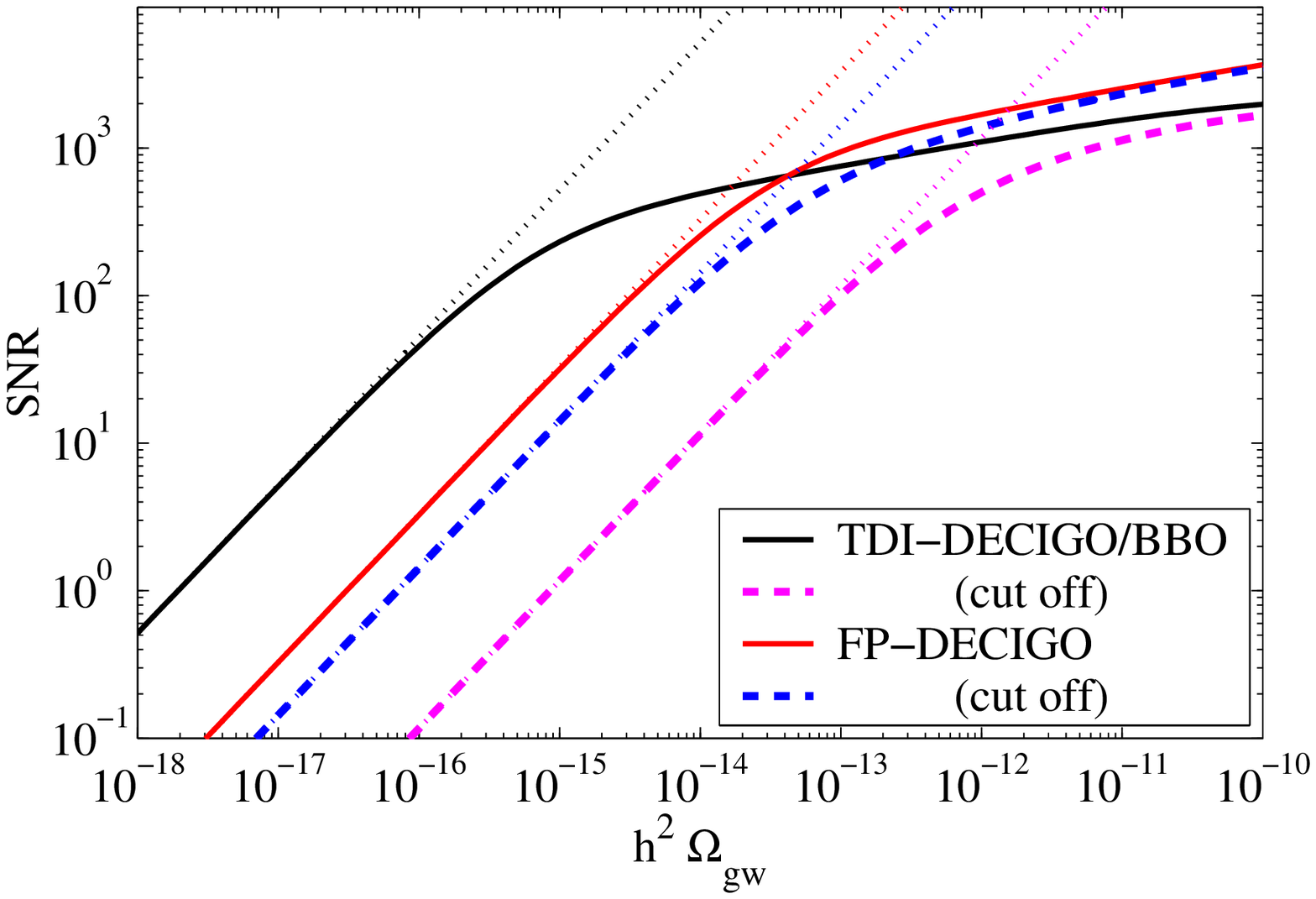}
\caption{
\label{fig:optimal_filter}
{\it Left}: optimal filter functions for isotropic GWB with various amplitudes $\ogw$ in the case of TDI-DECIGO. In plotting the functions, all the filters are normalized to have maximum magnitude equal to unity. {\it Right}: signal-to-noise ratio (SNR) as function of $h^2 \ogw$ with and without the low-frequency cutoff. 
Dashed curves shows the SNR with the cutoff frequency 
$f_{\mathrm{cut}}=0.2$ Hz.
Respective dotted thin lines depict the SNR based on the weak-signal approximation (\ref{eq:SNR_weak}).
}
\end{center}
\end{figure}

Clearly, the effect of low-frequency cutoff is significant for 
TDI-DECIGO/BBO. 
The minimum detectable amplitude for $f_{\rm cut}\gtrsim0.1$ Hz becomes $100$ times worse. 
By contrast, as long as the cutoff frequency is below $0.1$Hz, the minimum detectable amplitude of FP-DECIGO almost remains unchanged.
As a result, with the cutoff frequency $f_{\rm cut}=0.2$ Hz, the minimum amplitude $\Omega_{{\rm gw},0}h^2$ is changed to 
\begin{equation}
h^2\,\Omega_{{\rm gw},0}^{\mathrm{cutoff}} =\left\{ 
\begin{array}{cl}
5.8 \times 10^{-15} ~ 
{\displaystyle \left(\frac{\mathrm{SNR} }{5}\right)
\left(\frac{T_{\rm obs}}{1 \mathrm{year}}\right)^{-1/2} }
& (\mbox{TDI-DECIGO}) \\
4.8 \times 10^{-16} ~
 {\displaystyle \left(\frac{\mathrm{SNR}}{5}\right)
\left(\frac{T_{\rm obs}}{1 \mathrm{year}}\right)^{-1/2} }
& (\mbox{FP-DECIGO}) \\
1.2 \times 10^{-19}  ~
 {\displaystyle \left(\frac{\mathrm{SNR}}{5}\right)
\left(\frac{T_{\rm obs}}{1 \mathrm{year}}\right)^{-1/2} }
& (\mbox{ultimate DECIGO}) 
\end{array}
\right.
\end{equation}
Thus, in contrast to the previous estimate (\ref{eq:omega_no_cutoff}), 
FP-DECIGO has a potential to achieve better sensitivity than 
TDI-DECIGO/BBO, and the frequency covered by FP-DECIGO would be practically important to probe the primordial GWB. 
(See also Fig.~\ref{fig:optimal_filter} and discussions below.)
Nevertheless, it is premature to conclude that the slightly higher
frequency around $0.5-5$ Hz is the best observational window. 
Even above the frequency $f\sim0.2$ Hz, we still have foregrounds produced by the resolvable binaries made of neutron stars or black holes, which must be subtracted perfectly \cite{Cutler:2005qq}. 
Further, there might exist a large confusion noise arising from cosmological supernovae and/or hypothetical early population of massive stars \cite{Buonanno:2004tp}, which could dominate over the GWB of inflationary origin. 
These points are extremely important to give a practical mission design and must be clarified.

So far, we have quantified the SNR and the minimum detectable amplitude of $\ogw$ by taking the weak-signal limit. 
Before closing this subsection, we discuss the validity of the weak-signal approximation. The left panel of Fig.~\ref{fig:optimal_filter} shows the optimal filter functions $\widetilde{Q}(f)$ for TDI-DECIGO/BBO in cases with various amplitude of GWBs. 
Assuming the flat spectra $\ogw\propto f^0$, all the filter functions show a symmetric single-peak on a logarithmic scale of the frequency. 
As increasing the amplitude of $\ogw$, the location of the peak of the generalized optimal filter (\ref{eq:filter2}) [or Eq.(\ref{eq:filter_using_gamma})] moves to higher frequency,  while the filter function in the weak-signal approximation (\ref{eq:filer1}) remains unchanged (apart from the overall normalization).
In the right panel of Fig.\ref{fig:optimal_filter}, the SNRs for GWB are quantified taking account of the low-frequency cutoff.
The $\mathrm{SNR}$ taking the weak-signal approximation ({\it thin-dotted}) generally tends to overestimate the signal-to-noise estimated from general expression 
(\ref{eq:SNR_using_gamma}) ({\it thick}) and the discrepancy becomes significant above $\mathrm{SNR}\sim100$. 
This is because the general expression of $\mathrm{SNR}$ includes the GWB signal in both of the numerator and the denominator, while no such signal appears in the denominator when taking the weak-signal limit.  
Nevertheless, for sufficiently small amplitude of $\ogw$, the $\mathrm{SNR}$ of weak-signal approximation converges to the values obtained from the generalized optimal filter, as anticipated from the left panel.  
Although the convergence property depends on the cutoff frequency 
$f_{\rm cut}$ as well as the interferometer design, one can apply the weak-signal approximation to the correlation analysis of space interferometers as long as 
$\ogw\lesssim 10^{-15}$.

\subsection{Anisotropic case}
\label{subsec:anisotropic_GWB}

GWBs may exhibit anisotropic components in the sky distribution. 
This seems quite natural because the GWB of astrophysical origin can trace the spatial distribution of luminous galaxies, which gives a strong clustering pattern on small angular scales. 
Further, the primordial GWB generated during the inflation may also give an anisotropic component, like the CMB. 
In particular, a dipole moment arising from the proper motion of the solar-system barycenter would be observable with a future mission of space interferometer. 
Indeed, COBE \cite{Bennett:1996ce} and WMAP \cite{Bennett:2003bz} satellites have detected and determined the dipole moment of CMB, which is attributed to the motion of the solar system with respect to the CMB rest frame. The CMB dipole has amplitude $3.35$ mK towards the Galactic coordinates $(l',b')=(264^\circ, 48^\circ)$ or the ecliptic coordinates 
$(\phi_{\rm E}',\theta_{\rm E}')=(172^\circ, -11^\circ)$. 
This implies that our local system is moving with a velocity 
$\beta= v/c= 1.23\times10^{-3}$ in that direction, and then the observed frequency $f_{\mathrm{obs}}$ suffers from the Doppler shift.
If the CMB rest frame and the isotropic GWB rest frame are identical, the anisotropy induced by our proper motion towards 
$\mathbf{\Omega}_{\rm E}'=(\phi_{\rm E}',\theta_{\rm E}')$ becomes 
\begin{eqnarray}
    S_h (f_{\mathrm{obs}}) = S_h^{(0)}(f)\,
        \left[
        1 + 
        N \beta 
        \Bigl\{  \cos\theta_{\rm E} \cos \theta_{\rm E}' 
            \cos(\phi_{\rm E} - \phi_{\rm E}') + 
     \sin \theta_{\rm E} \sin \theta_{\rm E}'
            \Bigr\}
        \right].
\end{eqnarray}
Here 
$ S_h^{(0)} (f)$ denotes the isotropic component of intensity distribution and 
$N=\frac{ d\ln S_h}{d \ln f}$ is the tilt of the spectrum.
The induced multipole moments 
$p_{\ell m}^{\rm E}(f)$, which is defined by 
$S_h =\sum_{\ell m}p_{\ell m}^{\rm E} 
Y_{\ell m} $ in the ecliptic coordinates, are 
\begin{eqnarray}
    p^{\rm E}_{00} (f) = \sqrt{4\pi}\,\,  S_h^{(0)}, 
    \quad
    p^{\rm E}_{1 0}(f)= N \beta \, \sqrt{ \frac{8\pi}{3} } 
    \cos\theta'_{\rm E}\,\,S_h^{(0)},
    \quad
    p^{\rm E}_{1,\pm 1}(f) = \mp N \beta \sqrt{ \frac{2\pi}{3} }
    e^{\mp i \phi'_{\rm E}} \sin\theta'_{\rm E} ~ S_h^{(0)}.
\end{eqnarray}
Based on this, one finds that the non-vanishing components of the correlation signal $C_{{\scriptscriptstyle IJ},m}(f)$ given by 
Eq.~(\ref{eq:def_of_CIJ_m}) are $m=0,\,\pm1$ and $\pm2$: 
\begin{eqnarray}
C_{{\scriptscriptstyle IJ},0} (f) &=& \frac{1}{16\pi} 
\Bigl\{
    4 a_{00} \, p_{00}^{{\rm E}*} + a_{1,-1} \,p_{1,-1}^{{\rm E}*} + 
    2a_{10} \,p_{10}^{{\rm E}*}   + a_{11} \,p_{11}^{{\rm E}*} 
\Bigr\}, 
\cr
C_{{\scriptscriptstyle IJ},1} (f) &=& \frac{1}{8\pi} \sqrt{ \frac{3}{2}}
\Bigl\{
      a_{11} \, p_{10}^{{\rm E}*} - a_{10} \, p_{1,-1}^{{\rm E}*}  
\Bigr\}, 
\cr
C_{{\scriptscriptstyle IJ},2} (f) &=& \frac{3}{16\pi} a_{11} \, 
p_{1,-1}^{{\rm E}*} 
\label{eq:c_IJ, m=0,1,2}
\end{eqnarray}
and $   C_{{\scriptscriptstyle IJ}, m} (f) 
      = C_{{\scriptscriptstyle IJ},-m}^*(-f)$ for $I\neq J$. 
Here, the coefficients 
$a_{\ell m}(f)$ represent the multipole moments of the antenna pattern function $\mathcal{F}_{\scriptscriptstyle IJ}(f,\mathbf{\Omega})$ defined at the detector's rest frame. 
The relation between the multipole moments defined at the ecliptic frame and detector's rest frame is given by the Euler rotation matrix and an explicit expression can be written in terms of the Wigner $D$ matrices 
\cite{Kudoh:2004he,Taruya:2005yf,Cornish:2002bh,Seto:2004np}.

Since the $m=0$ component of the correlation signal is dominated by the isotopic component of GWB, only the $m=1$ and $2$ components are relevant for detecting the anisotropies of GWB. 
For TDI-DECIGO/BBO as well as ultimate DECIGO, we further note that the multipole coefficient $a_{10}$ vanishes for the cross-correlated $X$ variables of the star-like configuration. 
Hence the $m=1$ component of correlation signal contains only the information about 
$p_{10}^{\rm E}$. 
Assuming a flat spectrum of $\ogw$ in the observational band, our estimates of (\ref{eq:SNR2_weak}) for the future space interferometers are 
\begin{eqnarray}
&& (\mathrm{SNR})_1 
= 5 \, \left(\frac{T_{\mathrm{obs}}}{1\,\mathrm{year}}\right)^{1/2}
\times
\begin{cases}
\displaystyle
\frac{h^2 \ogw }{ 1.2 \times 10^{- 12}} 
\quad & (\text{TDI-DECIGO/BBO})
\vspace{0.2cm}
\\
\displaystyle 
\frac{h^2 \ogw }{ 2.0 \times 10^{- 10}} 
 \quad &(\text{FP-DECIGO})
\vspace{0.2cm}
\\
\displaystyle 
\frac{h^2 \ogw }{ 1.6 \times 10^{- 16}} 
 \quad &(\text{Ultimate DECIGO})
\end{cases}
\end{eqnarray}
and
\begin{eqnarray}
&& (\mathrm{SNR})_2 
= 5 \, \left(\frac{T_{\mathrm{obs}}}{1\,\mathrm{year}}\right)^{1/2}
\times
\begin{cases}
\displaystyle
   \frac{h^2 \ogw  }{ 5.3 \times 10^{-12}} 
   \quad &(\text{TDI-DECIGO/BBO})
\vspace{0.2cm}
\cr 
\displaystyle
   \frac{h^2 \ogw }{ 8.3 \times 10^{-10}} 
   \quad &(\text{FP-DECIGO}) 
\vspace{0.2cm}
\cr 
\displaystyle
   \frac{h^2 \ogw }{ 6.7 \times 10^{-16}} 
   \quad &(\text{Ultimate DECIGO} )
\end{cases}.
\end{eqnarray}
It is thus challenging problem to test observationally whether the CMB rest frame and the GWB rest frame are identical.  
If the amplitude of isotopic GWB is larger than the values listed above, we could observe the induced dipole moment of GWB and tackle the problem.  
We note that at the frequency below $0.2$ Hz, cosmological population of binaries constitutes the GWB with amplitude $\ogw \sim 10^{-11}$ 
\cite{Farmer:2003pa,Schneider:2000sg}.  
Therefore, the induced dipole moment of the astrophysical foreground would be detectable.  
This is analogous to the observation of a velocity dipole in the distribution of radio galaxies \cite{Blake:2002gx}.

Notice that the detectability of anisotropic components depends on not only the intensity distribution of GWB but also the angular sensitivity of antenna pattern function for space interferometer. 
To better understand the extent to which the next-generation space interferometers can probe the anisotropies of GWB, it may be helpful to quantify the strain sensitivity for each detectable multipole moment. 
Likewise Eq.~(\ref{eq:def_h_eff}), we introduce the effective strain sensitivity 
$h_{\rm eff}^{(\ell)}(f)$ for multipole moment $\ell$ 
\footnote{
We can give a theoretical basis for this effective strain sensitivity. 
A natural definition of effective strain sensitivity for each multipole moment is 
$ h_{\mathrm{eff}}^{(\ell)} = \sqrt{S_{h (\ell)}}$, 
where the effective power spectrum for each harmonics is defined by
$
 [S_{h (\ell)}]^2  
 \equiv 
 \sum_m  |p_{\ell m}^E|^2 / [ 4\pi (2\ell+1) ]
$.
Then the problem is how we define SNR for each multipole.
Since the antenna pattern function (in the detector's rest frame) and GW luminosity distribution can be expanded by the spherical harmonics, 
$C_{IJ,k}(f)$ is also given in terms of the multipole coefficients, like Eq.~(\ref{eq:c_IJ, m=0,1,2}).
Substituting such expansion into Eq.~(\ref{eq:SNR2_weak}), we can naturally introduce an appropriate SNR for each multipole moment. 
Then after some approximation, we arrive at the result (\ref{eq:def_h_eff_ell}).
}
\cite{Kudoh:2004he}:   
\begin{equation}
    h_{\rm eff}^{(\ell)}(f) = \overline{\mathrm{SNR}}^{1/2}\,
    \left\{4\pi\,
    \frac{N_{\scriptscriptstyle I}(f)N_{\scriptscriptstyle J}(f)}
    {\sigma_{\ell}^2(f)\,\Delta f\,\,T_{\rm obs}}\right\}^{1/4}.  
    \label{eq:def_h_eff_ell}
\end{equation}
Here, the quantity $\sigma_{\ell}(f)$ means the rotationally invariant representation of the angular power of antenna pattern function 
$\mathcal{F}_{\scriptscriptstyle IJ}(f;\mathbf{\Omega})$ for multipole moment $\ell$ \cite{Cornish:2001hg,Kudoh:2004he}: 
\begin{equation}
    \sigma^2_{\ell}(f)= \frac{1}{2\ell+1}\sum_{m=-\ell}^{\ell}
    \left|a_{\ell m}(f)\right|^2
\end{equation}
with $a_{\ell m}$ being the multipole moment of antenna pattern function at detector's rest frame. 
Note that we have $\sigma_{0}^2=4\pi\,|(2/5)\,\gamma_{\scriptscriptstyle IJ}|^2$. 
Thus, for the monopole moment $(\ell=0)$, the definition 
(\ref{eq:def_h_eff_ell}) correctly recovers the effective strain sensitivity for the isotropic GWB in (\ref{eq:def_h_eff}).

\begin{figure}[tb]
\begin{center}
\includegraphics[width=8.5cm,clip]{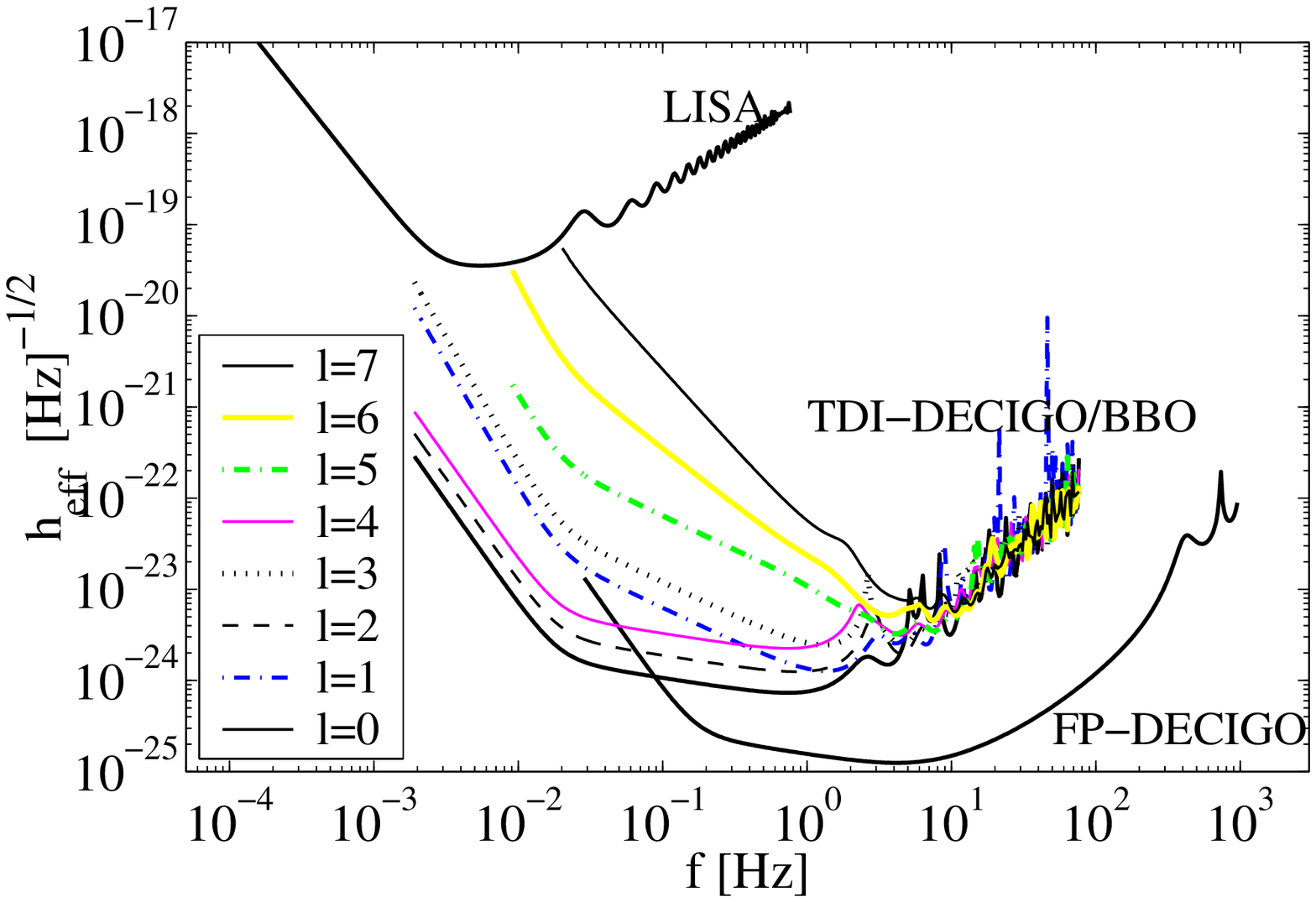}
\hspace*{0.5cm}
\includegraphics[width=8.5cm,clip]{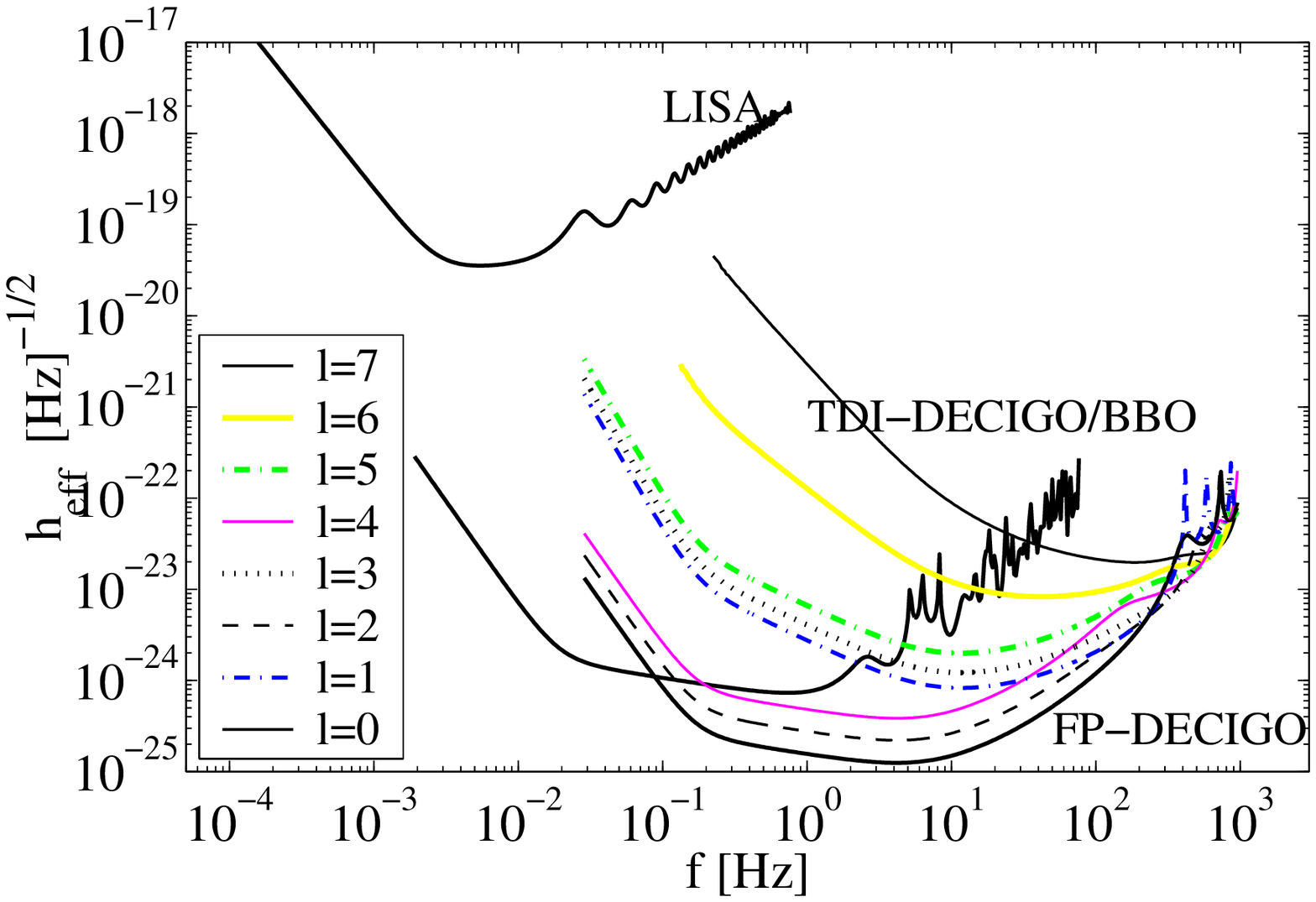}
\caption{
\label{fig:h_eff for ells}
Effective strain amplitude $h_{\mathrm{eff}}$ for TDI-DECIGO/BBO 
({\it left}) and FP-DECIGO ({\it right}).
In plotting the sensitivity curves for TDI-DECIGO/BBO, we specifically consider the cross-correlation between the TDI $X$-variables extracted from the nearest spacecrafts in the star-like configuration. 
In both cases, the interferometers are most sensitive to lower even multipoles of $\ell=0,2,4$. Because of the hexagonal form, the detectors are also sensitive to lower odd multipoles $\ell=1, 3, 5$.
The sensitivity to higher multipoles $\ell \ge 6$ is very poor and this fact is especially evident in the low frequency band. 
}
\end{center}
\end{figure}

Fig.~\ref{fig:h_eff for ells} shows the effective angular sensitivity of 
$\overline{\mathrm{SNR}}=5$ for the next-generation space interferometers TDI-DECIGO/BBO ({\it left}) and FP-DECIGO ({\it right}). 
Similar to Fig.~\ref{fig:h_eff for omega_GW}, we set the observation time $T_{\rm obs}$ to $1$ year and the band width $\Delta f$ to $f/10$. 
Interestingly, the resultant strain sensitivity for each multipole moment shows a band-like structure.
The future space interferometers are more sensitive to the even multipoles $\ell=0,\,2$ and $4$ than the odd multipoles $\ell=1,\,3$ and $5$. These behaviors can be ascribed to both the low-frequency properties of antenna pattern function and the geometric configuration of space interferometers \cite{Kudoh:2004he}. 
As a consequence, the angular resolution of space interferometers is rather poor and the detectable multipole moments are very restrictive. 
At $f\sim10$Hz, the sensitivity of FP-DECIGO may reach $\ell\sim5$ with the effective sensitivity $h_{\rm eff}^{(\ell)}\simeq 5\times10^{-24}$ 
Hz$^{-1/2}$, corresponding to $h^2\ogw\sim 10^{-9}$. 
At the same frequency band, the angular sensitivity of TDI-DECIGO/BBO is much worse, as well as the response to the GWB becomes quite complicated because of the oscillatory nature in the antenna pattern function.

\section{Summary}
\label{sec:summary}

We are currently in the early stage to make a conceptual design of next-generation space interferometers.
The planned future missions will be dedicated to detect the stochastic GWB of cosmological origin. 
In this paper, we have discussed the detection of such GWB via correlation analysis and studied prospects for direct measurement of both isotropic and anisotropic components of GWBs by future missions. For this purpose, we have presented the general expressions for signal-to-noise ratio. 
Taking the weak-signal limit, the optimal filter functions were defined so as to increase the sensitivity to the GWB signal. 
In the isotropic case, a generalized optimal filter was derived, which can be used in any combination of output signals with arbitrary large amplitude of GWB signals.

Based on this formalism, we have also demonstrated the feasibility of
proposed future missions to detect the GWB produced during the inflation. 
Due to the geometric properties of spacecraft configuration, 
LISA would not probe isotopic GWBs by the cross-correlation analysis and the accessible minimum value of $\ogw$ is severely restricted by the detector's intrinsic noise \cite{Kudoh:2004he}. 
For the flat spectrum of $\ogw$, the minimum detectable value could reach at most about $h^2 \ogw \gtrsim10^{-11}$. 
On the other hand, the space interferometers, like TDI-DECIGO/BBO, which form a star-like configuration of spacecrafts will improve this limit greatly by seven orders of magnitude, i.e.,  $h^2 \ogw \lesssim 10^{-18}$, which is almost comparable to the sensitivities expected for future experiments of CMB polarization (e.g.,~\cite{Smith:2005mm}). 
However, there might possibly exist several astrophysical foregrounds in the observed frequency band 
\cite{Farmer:2003pa,Schneider:2000sg,Buonanno:2004tp}, which act as a disturbance of detecting the primordial GWB. 
We have examined the effect of foreground sources by introducing the cutoff frequency and found that TDI-DECIGO/BBO is quite sensitive to the low-frequency cutoff, while the sensitivity of FP-DECIGO to the primordial GWB almost remains unchanged, resulting in the detection level $h^2 \ogw\sim10^{-16}$. 
Although there still remain some problems concerning the point-source subtraction, the result indicates that FP-DECIGO is a potentially suited design for detecting the primordial GWB.

In addition to the detectability of isotropic GWBs, we have investigated the directional sensitivity of next-generation space interferometer to the anisotropic GWB. As a demonstration, the dipole anisotropy induced by proper motion of local observer was considered. 
For cross-correlation signals extracted from the star-like spacecraft configuration, the interferometers are more sensitive to the even modes ($\ell=0,2, 4)$ than the odd modes $(\ell=1,3,5)$ as anticipated from Ref.~\cite{Kudoh:2004he}.  
Accordingly, the detection of dipole anisotropy would be possible only when the isotropic component is $h^2 \ogw\gtrsim10^{-11}$. 
Hence, although very interesting, it would be hard to probe whether the CMB rest frame and a GWB rest frame are both identical or not.

In any case, space interferometers will be a cornerstone for a new understanding of the Universe. We hope that the present study will be helpful for developing and fixing the pre-conceptual design of next-generation of space interferometers.

\acknowledgments
We would like to thank Masaki Ando for his helpful advice on the future space-interferometers. We also thank Naoki Seto and the DECIGO working group for useful comments and discussions. 
H.K is grateful to Takahiro Tanaka for his helpful comments. 
This work was partially supported by JSPS (Japan Society for the 
Promotion of Science) fellowship (H.K, Y.H and T.H). 

    \appendix

\section{Derivation of expectation values for 
output signals}
\label{appendix:derivation}

In this appendix, based on the statistical assumption in 
Sec.\ref{sec:preliminaries}, we derive the expectation values for output signals obtained from the gravitational-wave detectors.

The Fourier component of the output signal $s_{\scriptscriptstyle I}(t)$ is
\begin{equation}
\widetilde{s}_{\scriptscriptstyle I}(f,t)=\int_{t-\tau/2}^{t+\tau/2}dt'\,
s_{\scriptscriptstyle I}(t') \,e^{i\,2\pi\,f\,t'}.
\end{equation}
Then the signal defined in Eq.(\ref{eq:correlation}) becomes
\begin{equation}
    S_{\scriptscriptstyle IJ}(t)= 
\int_{-\infty}^{\infty} df 
df'
df'' \,\,
\widetilde{s}^*_{\scriptscriptstyle I}(f,t)
\widetilde{s}_{\scriptscriptstyle J}(f',t)
\widetilde{Q}(f'')\,\delta_{\tau}(f-f'')\delta_{\tau}(f''-f')\,\,
e^{i\,2\pi\,(f-f')t},
\label{eq:correlation_Fourier}
\end{equation}
where the quantity $\delta_{\tau}(f)$ represents the finite-time approximation to the Dirac delta function $\delta(f)$ 
\cite{Allen:1997gp,Allen:1997ad}: 
\begin{equation}
\delta_{\tau}(f)\equiv \int_{-\tau/2}^{\tau/2}dt'\,e^{i\,2\pi\,f\,t'}
= \frac{\sin{(\pi\,f\,\tau)}}{\pi\,f},
\label{eq:def_delta_tau}
\end{equation}
which reduces to $\delta(f)$ in the limit $\tau\to \infty$, but has the property $\delta_\tau(0)=\tau$.
Based on the expression (\ref{eq:correlation_Fourier}), our task is to calculate the averaged quantities $\mu_{\scriptscriptstyle IJ}$ and $\mu_{{\scriptscriptstyle IJ},m}$ and their variances $\Sigma_{{\scriptscriptstyle IJ}}^2$ and $\Sigma_{{\scriptscriptstyle IJ},m}^2$. 
These are separately calculated in Appendix \ref{subsec: mu} and 
\ref{subsec: sigma}. 
%
%
%
%
%
%
%
%
\subsection{$\mu_{\scriptscriptstyle IJ}$ and $\mu_{{\scriptscriptstyle IJ},m}$}
\label{subsec: mu}
%
%
%
%
%
%
%
%
%
%
In this case, it is sufficient to evaluate the quantity $\langle S_{\scriptscriptstyle IJ}(t)\rangle$. 
The ensemble average 
$\langle \widetilde{s}^*_{\scriptscriptstyle I}(f,t)
\widetilde{s}_{\scriptscriptstyle J}(f',t)\rangle$ is divided into two parts:
\begin{eqnarray}
\langle \widetilde{s}^*_{\scriptscriptstyle I}(f,t)
\widetilde{s}_{\scriptscriptstyle J}(f',t)\rangle
=\langle
\widetilde{h}^*_{\scriptscriptstyle I}(f,t)
\widetilde{h}_{\scriptscriptstyle J}(f,t)\rangle + 
\langle \widetilde{n}_{\scriptscriptstyle I}^*(f) 
\widetilde{n}_{\scriptscriptstyle J}(f')\rangle. 
\nonumber 
\end{eqnarray}
Using the definitions (\ref{eq:def_GWBspec}) and (\ref{eq:def_noise}) as well as the relation (\ref{eq:h_I}), we obtain 
\begin{eqnarray}
\langle \widetilde{s}^*_{\scriptscriptstyle I}(f,t)
\widetilde{s}_{\scriptscriptstyle J}(f',t)\rangle
&=& \frac{1}{2}\,\delta(f-f') \,\left[
\int \frac{d\mathbf{\Omega}}{4\pi}\,
\mathcal{F}_{\scriptscriptstyle IJ}(f,\mathbf{\Omega};t,t)\,
S_h(f,\mathbf{\Omega}) + \delta_{\scriptscriptstyle IJ}\,
N_{\scriptscriptstyle I}(f)
\right],
\nonumber
\end{eqnarray}
where we defined
\begin{equation}
\mathcal{F}_{\scriptscriptstyle IJ}(f,\mathbf{\Omega};t,t')\equiv
e^{i\,2\pi\,f\,\mathbf{\Omega}\left[x_{\scriptscriptstyle I}(t)- 
x_{\scriptscriptstyle J}(t')\right]}
\sum_{A=+,\times}
\left[ D^{ab\,*}_{\scriptscriptstyle I}(\mathbf{\Omega},f;t)\,
{\rm e}^{A}_{ab}(\mathbf{\Omega}) \right]\,\,
\left[ D^{cd}_{\scriptscriptstyle J}(\mathbf{\Omega},f;t')\,
{\rm e}^{A}_{cd}(\mathbf{\Omega}) \right].
\label{eq:def_antenna_pattern}
\end{equation}
If $t=t'$, this is the so-called antenna pattern function, which is related with a overlap reduction function [see Eq.(\ref{eq:overlap_reduc_func})]. 
Substituting the above equation into 
$\langle S_{\scriptscriptstyle IJ}(t)\rangle$, we have 
\begin{equation}
\mu_{\scriptscriptstyle IJ}(t)=\langle S_{\scriptscriptstyle IJ}(t)\rangle =
\int_{-\infty}^{\infty} df 
 df'' \,\,
\widetilde{Q}(f'')\,\left[\delta_{\tau}(f-f'')\right]^2\,
\frac{1}{2}\left[
\int \frac{d\mathbf{\Omega}}{4\pi}\,
\mathcal{F}_{\scriptscriptstyle IJ}(f,\mathbf{\Omega};t,t)\,
S_h(f,\mathbf{\Omega}) + \delta_{\scriptscriptstyle IJ}\,
N_{\scriptscriptstyle I}(f)
\right].
\end{equation}
For our interest of the observed frequency $f\gg 1/\tau$, the function $\delta_{\tau}(f-f'')$ is sharply peaked around 
$f\sim f''$ and one can replace one of the finite-time delta functions by an ordinary Dirac function. Then the above equation is reduced to 
\begin{equation}
\mu_{\scriptscriptstyle IJ}(t) 
\approx 
\tau\,
\int_{-\infty}^{\infty} \frac{df}{2}\, \widetilde{Q} 
\left[
\int \frac{d\mathbf{\Omega}}{4\pi} 
\mathcal{F}_{\scriptscriptstyle IJ}  
S_h  + \delta_{\scriptscriptstyle IJ}\,
N_{\scriptscriptstyle I} 
\right],
\label{appendix:corr_C}
\end{equation}
where we have used the fact that $\delta_{\tau}(0)=\tau$.  
Note that the above result does not assume a statistical isotropy of 
GWB and can generally apply to an anisotropic case of GWB.

Now assuming the isotropy of GWB and using the definition 
(\ref{eq:corr_C}), this simply leads to the final result (\ref{eq:mu}). 
Also, it is easy to derive $\mu_{{\scriptscriptstyle IJ},m}$. 
With $\omega=2\pi/T_{\rm orbit}$, we have 
\begin{equation}
\mu_{{\scriptscriptstyle IJ},m} = \frac{1}{T_{\rm orbit}}
\int_0^{T_{\rm orbit}} dt\,\,e^{-i\,m\,\omega\,t} 
\langle S_{\scriptscriptstyle IJ}(t)\rangle, 
\end{equation}
Thus, substituting Eq.~(\ref{appendix:corr_C}) into the above equation, with a help of definition (\ref{eq:def_of_CIJ_m}), we finally obtain 
Eq.~(\ref{eq:mu_m}).

\subsection{$\Sigma_{\scriptscriptstyle IJ}^2$ and 
$\Sigma_{{\scriptscriptstyle IJ},m}^2$}
\label{subsec: sigma}

The derivation of $\Sigma_{\scriptscriptstyle IJ}^2$ and $\Sigma_{{\scriptscriptstyle IJ},m}^2$ becomes slightly complicated. 
We first write down the second-order correlation 
$\langle S{\scriptscriptstyle IJ}(t) S{\scriptscriptstyle IJ}(t')\rangle$: 
\begin{eqnarray}
&&\langle S{\scriptscriptstyle IJ}(t) S{\scriptscriptstyle IJ}(t')\rangle =
\int_{-\infty}^{\infty}df_1 \,
df_2 \,
df_3 \,
df_4 \,
df_5 \,
df_6 \,\,
\widetilde{Q}(f_3)\,\widetilde{Q}(f_6)\,
\,e^{i\,2\pi\{(f_1-f_2)t+(f_4-f_5)t'\}}
\nonumber \\
&& \quad\quad\quad\quad~~~~\times\,
\delta_{\tau}(f_1-f_3)\delta_{\tau}(f_3-f_2)
\delta_{\tau}(f_4-f_6)\delta_{\tau}(f_6-f_5)\,
\bigl\langle 
\widetilde{s}_{\scriptscriptstyle I}^*(f_1,t)
\widetilde{s}_{\scriptscriptstyle J}(f_2,t)
\widetilde{s}_{\scriptscriptstyle I}^*(f_4,t')
\widetilde{s}_{\scriptscriptstyle J}(f_5,t')
\bigr\rangle. 
\label{eq:ensemble_SS'}
\end{eqnarray}
The above equation includes a complicated ensemble average 
$\left\langle 
\widetilde{s}_{\scriptscriptstyle I}^*(f_1,t)
\widetilde{s}_{\scriptscriptstyle J}(f_2,t)
\widetilde{s}_{\scriptscriptstyle I}^*(f_4,t')
\widetilde{s}_{\scriptscriptstyle J}(f_5,t')
\right\rangle$. Assuming the Gaussianity of the output data 
$s_{\scriptscriptstyle I}(f,t)$ and no statistical correlations between signal and noise, this quantity is divided into the 12 terms by means of the Wick theorem. 
Collecting these terms and repeating the similar calculation in each term as done in Appendix \ref{subsec: mu}, a lengthy but straightforward calculation yields
\begin{eqnarray}
\left\langle S_{\scriptscriptstyle IJ}(t)
S_{\scriptscriptstyle IJ}(t') \right\rangle
&=&
\tau^2 \
\left\{
\int_{-\infty}^{\infty}\frac{df}{2}\widetilde{Q} 
\Bigl[C_{\scriptscriptstyle IJ}(f;t,t) 
+ \delta_{\scriptscriptstyle IJ} N_{\scriptscriptstyle I} \Bigr]
\right\}
\left\{
\int_{-\infty}^{\infty}\frac{df}{2}\widetilde{Q} 
\Bigl[C_{\scriptscriptstyle IJ}(f;t',t')  
+ \delta_{\scriptscriptstyle IJ} N_{\scriptscriptstyle I} \Bigr]
\right\}
\cr
&+&
\,\frac{\tau}{2}\,\int_{-\infty}^{\infty}\frac{df}{2}\,\widetilde{Q}^2 
\left[\left\{C_{\scriptscriptstyle IJ}(f;t,t')\right\}^2 +
\delta_{\scriptscriptstyle IJ} 
N_{\scriptscriptstyle I} \Bigl\{ 
 N_{\scriptscriptstyle I}   + 
 C_{\scriptscriptstyle II}(f;t,t)+C_{\scriptscriptstyle II}(f;t',t') 
 \Bigr\}
\right]
\cr
&+&
\,\frac{\tau}{2}\,\int_{-\infty}^{\infty}
\frac{df}{2}\,\left|\widetilde{Q} \right|^2
\Bigl[
C_{\scriptscriptstyle II}(f;t,t')C_{\scriptscriptstyle JJ}(f;t,t')+
C_{\scriptscriptstyle II}(f;t,t')N_{\scriptscriptstyle J} +
C_{\scriptscriptstyle JJ}(f;t,t')N_{\scriptscriptstyle I} +
N_{\scriptscriptstyle I} N_{\scriptscriptstyle J} 
\Bigr]. 
\nonumber
\end{eqnarray}
Here, we have used the same notation $C_{\scriptscriptstyle IJ}(f;t,t')$ as defined in (\ref{eq:corr_C}) to express the statistical quantity: 
\begin{equation}
C_{\scriptscriptstyle IJ} (f;t,t')= \int \frac{d\mathbf{\Omega}}{4\pi}\,
S_h(f,\mathbf{\Omega})\, 
\mathcal{F}_{\scriptscriptstyle IJ}(f,\mathbf{\Omega};t, t'). 
\label{eq:corr_C2} 
\end{equation}

Now, we define 
\begin{eqnarray}
\Sigma_{\scriptscriptstyle IJ}^2(t,t')
\equiv
\left\langle 
    S_{\scriptscriptstyle IJ}(t)S_{\scriptscriptstyle IJ} (t') 
\right\rangle 
-
\left\langle S_{\scriptscriptstyle IJ} (t)
\right\rangle \left\langle S_{\scriptscriptstyle IJ}(t') \right\rangle,
\label{eq:def_Sigma_IJ}
\end{eqnarray}
which gives 
\begin{eqnarray}
&&\Sigma_{\scriptscriptstyle IJ}^2(t,t')=
\frac{\tau}{2}\,\int_{-\infty}^{\infty}\frac{df}{2}\,\widetilde{Q}^2 
\left[\left\{C_{\scriptscriptstyle IJ}(f;t,t')\right\}^2 +
\delta_{\scriptscriptstyle IJ}\,\Bigl\{ 
    N_{\scriptscriptstyle I} ^2 
+ [ C_{\scriptscriptstyle II}(f;t,t)+C_{\scriptscriptstyle II}(f;t',t')]
    N_{\scriptscriptstyle I} 
\Bigr\}
\right]
\cr
&&\quad\quad+
\,\frac{\tau}{2}\,\int_{-\infty}^{\infty}
\frac{df}{2}\,\left|\widetilde{Q} \right|^2
\Bigl[
  C_{\scriptscriptstyle II}(f;t,t')C_{\scriptscriptstyle JJ}(f;t,t')
+ C_{\scriptscriptstyle II}(f;t,t')N_{\scriptscriptstyle J} 
+ C_{\scriptscriptstyle JJ}(f;t,t')N_{\scriptscriptstyle I} 
+ N_{\scriptscriptstyle I} N_{\scriptscriptstyle J} 
\Bigr].
\label{eq:Sigm(t,t')}
\end{eqnarray}
Equating $t$ with $t'$ and assuming the isotropy of GWB, we obtain the final expression for $\Sigma_{\scriptscriptstyle IJ}^2$ given by Eq.~(\ref{eq:sigma2_general}).

On the other hand, the mean value 
$\Sigma^2_{{\scriptscriptstyle IJ},m}$ given by (\ref{eq:def_SNR2}) is obtained by computing the following quantity:
\begin{eqnarray}
\Sigma_{{\scriptscriptstyle IJ},mm'}^2 = 
\frac{1}{T_{\rm orbit}^2}
\int_0^{T_{\rm orbit}} dt \int_0^{T_{\rm orbit}} dt'\,\,
\Sigma_{\scriptscriptstyle IJ}^2(t,t')
\,e^{i\,\omega (mt-m't')}. 
\label{eq:Sigm(m,m')}
\end{eqnarray}
When $m=m'$, this coincides with the quantity 
$\Sigma^2_{{\scriptscriptstyle IJ},m}$. 
The evaluation of Eq.(\ref{eq:Sigm(m,m')}) seems rather difficult because the terms in Eq.(\ref{eq:Sigm(t,t')}) associated with the gravitational-wave signals involve the different-time correlation such as $C_{\scriptscriptstyle IJ}(f;t,t')$, which becomes non-vanishing even if $t\ne t'$. In practice, however, the effect of the different-time correlation becomes negligible if we consider the time-scales larger than 
$T_*$. The characteristic time $T_*$ is roughly evaluated as 
$T_*\simeq (2\pi\,\dot{x}\,f)$, where $\dot{x}$ means velocity of space craft. 
Thus, as long as the local observation time $\tau$ is chosen as 
$\tau\gg T_*$, one can approximate $C_{\scriptscriptstyle IJ}(f;t,t')$ as 
\begin{equation}
C_{\scriptscriptstyle IJ}(f;t,t')\approx T_*\,\,
\delta(t-t')\,C_{\scriptscriptstyle IJ}(f;t,t). 
\label{eq:approx}
\end{equation}
In Appendix \ref{appendix:details_of_calc}, we discuss the validity of this treatment in some details.

The approximation (\ref{eq:approx}) greatly simplifies the evaluation of 
(\ref{eq:Sigm(m,m')}). In addition to this, a careful treatment is necessary when we evaluate the terms consisting of the noise spectra only [see Eq.(\ref{eq:Sigm(t,t')})]. 
Since these terms have no explicit time-dependence, a naive calculation based on the expression (\ref{eq:Sigm(t,t')}) incorrectly drops their contribution to the variance $\Sigma_{{\scriptscriptstyle IJ},mm'}$.
In Appendix \ref{appendix:details_of_calc}, some tricks to evaluate these terms are also presented (see also Sec.IX of Ref.\cite{Allen:1997gp}). Taking into account of these treatments, the quantity $\Sigma_{{\scriptscriptstyle IJ},mm'}^2$ is finally reduced to 
\begin{eqnarray}
\Sigma_{{\scriptscriptstyle IJ},mm'}&\simeq&
\frac{\tau}{2}\left(\frac{T_*}{T_{\rm orbit}}\right)\,
\int_{-\infty}^{\infty}\frac{df}{2}\left\{ \widetilde{Q}^2(f)\,V_{mm'}(f) 
+ \left|\widetilde{Q}(f)\right|^2\,W_{mm'}(f)\right\}
\label{eq:result_Sigm(m,m')}
\end{eqnarray}
with the functions $V_{mm'}(f)$ and $W_{mm'}(f)$ being 
\begin{eqnarray}
V_{mm'}(f)
&=& \sum_n C_{{\scriptscriptstyle IJ},n-m}(f)
C_{{\scriptscriptstyle IJ},m'-n}(f) 
+
\delta{\scriptscriptstyle IJ}N_{\scriptscriptstyle I}
\left[ 2 C_{{\scriptscriptstyle II},m'-m}(f)  
+ 
\delta_{mm'}\,
\left(\frac{\tau}{T_*}\right)^2 N_{\scriptscriptstyle I} \right],
\nonumber
\\
W_{mm'}(f)
&=&  \sum_n C_{{\scriptscriptstyle II},n-m}(f)
C_{{\scriptscriptstyle JJ},m'-n}(f) + 
C_{{\scriptscriptstyle II},m'-m}(f) N_{\scriptscriptstyle J} + 
C_{{\scriptscriptstyle JJ},m'-m}(f) N_{\scriptscriptstyle I} + 
\delta_{mm'}\,\left(\frac{\tau}{T_*}\right)\,
N_{\scriptscriptstyle I} N_{\scriptscriptstyle J} ,
\nonumber
\end{eqnarray}
which leads to the final result 
(\ref{eq:sigma2_mu_general}) when taking $m=m'$.

\section{Antenna pattern functions and 
instrumental noises}
\label{appendix:antenna}

For the detection of GWB via correlation analysis, a crucial task is a choice of output signals because it affects the sensitivity to the stochastic GWBs through the antenna pattern function. 
In the main text, the correlation analysis was performed using the Michelson-type TDI signals called $X$ $(Y, Z)$ variables in the cases of TDI-DECIGO/BBO and ultimate DECIGO. 
For FP-DECIGO, the Fabry-Perot type interferometric variables were used. Here, we give a specific functional form of the antenna pattern functions used in the main text.

Let us recall the definition of antenna pattern function:
\begin{eqnarray}
\mathcal{F}_{\scriptscriptstyle IJ}(f,\mathbf{\Omega};t,t')\equiv
e^{i\,2\pi\,f\,\mathbf{\Omega}\left[x_{\scriptscriptstyle I}(t)- 
x_{\scriptscriptstyle J}(t')\right]}
\sum_{A=+,\times}
\left[ D^{ab\,*}_{\scriptscriptstyle I}(\mathbf{\Omega},f;t)\,
{\rm e}^{A}_{ab}(\mathbf{\Omega}) \right]\,\,
\left[ D^{cd}_{\scriptscriptstyle J}(\mathbf{\Omega},f;t')\,
{\rm e}^{A}_{cd}(\mathbf{\Omega}) \right].
\nonumber
\end{eqnarray}
Since we are especially concerned with the cross-correlation analysis between the signals extracted from the spacecrafts forming the star-like configuration (see Fig.~\ref{fig:hexagonal interferometer}), the position vectors $x_{\scriptscriptstyle I}$ and 
$x_{\scriptscriptstyle J}$ should be set to the nearest-neighbor vertices in each triangular configuration. 
Based on these vertices as starting points, let us denote the unit vectors pointing to the other spacecrafts forming the triangular configuration by $\mathbf{a}$ and $\mathbf{c}$ (and $\mathbf{a}'$ and $\mathbf{c}'$)
\footnote{These unit vectors inevitably depend on time, due to the orbital motion of spacecrafts. }. 
Then, the detector tensor for the TDI $X$-variable in the equal arm length limit becomes 
\begin{eqnarray}
D^{ij}_{\rm X}(\mathbf{\Omega},\hat{f};t) =\frac{1}{4}
\left(1-e^{-i\,2\hat{f}}\right)\,\,
\Bigl[
 \mathbf{a}^i(t) \otimes  \mathbf{a}^j(t) ~
\mathcal{T}\left(\mathbf{a}(t)\cdot\mathbf{\Omega},\,\hat{f}\right)
-
 {\mathbf{c}}^i (t) \otimes {\mathbf{c}}^j(t) ~
\mathcal{T}\left(-\mathbf{c}(t)\cdot\mathbf{\Omega},\,\hat{f}\right)
\Bigr], 
\label{eq:detector_tensor_X}
\end{eqnarray}
where $\mathcal{T}(\mathbf{u}\cdot\mathbf{\Omega},\hat{f})$ is the transfer function given by 
\begin{eqnarray}
\mathcal{T}(\mathbf{u}\cdot\mathbf{\Omega},\hat{f}) = 
e^{-i\,\hat{f}}\left\{
   \mbox{sinc}\left(\frac{\hat{f}(1-\mathbf{u}\cdot\mathbf{\Omega})}{2}\right) 
 e^{-(i/2)\hat{f}(1+\mathbf{u}\cdot\mathbf{\Omega})}
+ 
 \mbox{sinc}\left(\frac{\hat{f}(1+\mathbf{u}\cdot\mathbf{\Omega})}
{2}\right)\,\,e^{-(i/2)\hat{f}(-1+\mathbf{u}\cdot\mathbf{\Omega})}
\right\}.
\nonumber
\end{eqnarray}
Here $\hat{f}$ is the frequency normalized by the characteristic frequency  $f_*=(c/2\pi L)$, i.e. $\hat{f}=f/f_*$.
On the other hand, the detector tensor for Fabry-Perot signal becomes
\begin{equation}
D^{ij}_{\rm FP}(\mathbf{\Omega},f;t) =
\frac{1}{2} \Bigl[ 
  {\mathbf{a}}^i(t) \otimes {\mathbf{a}}^j (t) 
- {\mathbf{c}}^i(t) \otimes {\mathbf{c}}^j (t)  \Bigr]
\label{eq:detector_tensor_FP}
\end{equation}
Likewise, the detector tensor for another triangular configuration is obtained from 
(\ref{eq:detector_tensor_X}) or (\ref{eq:detector_tensor_FP}) just replacing the unit vectors  
$\mathbf{a}$ and $\mathbf{c}$ with $\mathbf{a}'$ and $\mathbf{c}'$.

It is important to notice that the spectral density of instrumental noises also depends on the choice of the interferometric signals. 
In the case of the TDI $X$-variables, one has 
\begin{equation}
N_{\rm X}(f) = 16\,\sin^2\hat{f}\,\left\{ 
N_{\rm shot}(f) + 2\left(1+\cos^2\hat{f} \right) N_{\rm accel}(f)
\right\}, 
\end{equation}
where $N_{\rm shot}=(S_{\rm shot}/L)^2$ and $N_{\rm accel} =(S_{\rm accel}/L)^2 (2\pi f)^{-4}$ represent the spectral density of shot noise and the acceleration noise, respectively.
The instrumental parameters $S_{\rm shot}$ and $S_{\rm accel}$ are listed in 
Table \ref{table: instrument parameter}.
The noise spectral density of Fabry-Perot signal is
\begin{equation}
N_{\rm FP}(f)= 
  N_\mathrm{shot}  
+ N_\mathrm{accel} 
+ N_\mathrm{rad}  ,
\end{equation}
where $N_\mathrm{rad}=S_\mathrm{rad}^2$ is the radiation pressure noise given in Table \ref{table: instrument parameter}.

\section{Some details on the derivation of  
Eq.(\ref{eq:result_Sigm(m,m')})}
\label{appendix:details_of_calc}

In this appendix, we give some detailed discussions on the derivation of the analytic expression (\ref{eq:result_Sigm(m,m')}).

First, we consider the validity of the approximation (\ref{eq:approx}). 
Since the time dependence of $C_{\scriptscriptstyle IJ}(f;t,t')$ arises only from the antenna pattern function (\ref{eq:def_antenna_pattern}) and the significance of the different-time correlation mainly comes from the phase factor in the antenna pattern function, i.e.,  
$e^{i\,2\,\pi\,f\,\mathbf{\Omega}
[x_{\scriptscriptstyle I}(t)-x_{\scriptscriptstyle I}(t')]}$, it would be better to focus on the role of the phase factor.  
Effectively, this term represents the coherence of the gravitational waves observed at the different detector sites and with the different times,  $x_{\scriptscriptstyle I}(t)$ and 
$x_{\scriptscriptstyle I}(t')$. For a given frequency $f$, the phase factor becomes a rapidly oscillating function of $\mathbf{\Omega}$ when the distance between the two detectors is sufficiently long, i.e.,  
$|x_{\scriptscriptstyle I}-x_{\scriptscriptstyle J}|\gg (2\pi\,f)^{-1}$.  
In this case, the phase factor almost cancels after integrating over the whole sky. The important notice is that the phase cancellation even occurs when $I=J$ (self-correlation case).  
This is because $x_{\scriptscriptstyle I}(t)\ne x_{\scriptscriptstyle I}(t')$ due to the motion of the detectors. Thus, for a sufficiently longer time-interval 
$|t-t'|\gg T_*=(2\pi\dot{x}_{\scriptscriptstyle I}f)^{-1}$, one expects that the correlation between the two different-time $t$ and $t'$ decays rapidly and the quantity 
$C_{\scriptscriptstyle IJ}(f;t,t')$ gives no meaningful information.

\begin{figure}[bt]
\begin{center}
\includegraphics[width=8cm,clip]{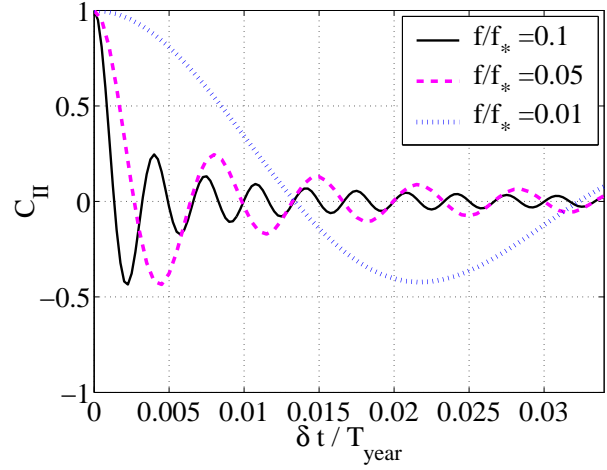}
\caption{
\label{fig:overlap}
Plot of $C_{\scriptscriptstyle IJ}(f;t,t+\delta t)$ with respect to the time difference $\delta t$ for TDI-DECIGO/BBO (self-correlation case).
The antenna pattern function is evaluated in the low frequency regime ($ f/f_* =  2\pi L f/c \ll 1 $), assuming flat isotropic GW spectrum ($S_h = \mathrm{const.})$. 
Since the amplitude of $C_{\scriptscriptstyle IJ}$ depends on $f$, we have appropriately normalized it for comparison. 
Due to the phase (overlap reduction) factor 
$e^{ i\, 2\pi f {\bf \Omega \cdot}[ \textbf{x}_I (t) -\textbf{x}_J(t') ] }$ in Eq.~(\ref{eq:def_antenna_pattern}), the oscillating function 
$C_{\scriptscriptstyle IJ}$ decays very rapidly as the relative time difference increases. 
The characteristic time scale of the decay is readily estimated as 
$\delta t \sim c (L/\dot x_I)/(2\pi L f) \sim 10^{-3} T_{\mathrm{year}} \times (0.1 f_*/f)$. 
}
\end{center}
\end{figure}

To show this explicitly, specifically using the optimal TDI $A$-variable 
\cite{Prince:2002hp}, we evaluate the quantity $C_{\scriptscriptstyle IJ}(f;t,t')$ in the self-correlation case.  
Figure~\ref{fig:overlap} shows the plot of $C_{\scriptscriptstyle IJ}(f;t,t+\delta t)$ as function of the relative time-difference $\delta t$, assuming a white signal ($S_h=$const.). 
As is expected, $C_{\scriptscriptstyle IJ}(f;t,t+\delta t)$ is sharply peaked around $\delta t\lesssim T_*$ and rapidly damped at $\delta t\gtrsim T_*$.  
Thus, as long as the timescale larger than the characteristic time $T_*$ is concerned, the approximation 
(\ref{eq:approx}) is valid. 
Note that in the correlation analysis for the anisotropic GWB, local observation time $\tau$ should be typically chosen as few weeks because of the low angular sensitivity to the stochastic GWB. 
Therefore, we have $\tau\gtrsim T_*$, and the validity of the approximation (\ref{eq:approx}) is always satisfied.

Let us next discuss somewhat tricky treatment to derive the final result 
(\ref{eq:result_Sigm(m,m')}). 
As we mentioned in Appendix \ref{subsec: sigma}, a naive substitution of the expression (\ref{eq:Sigm(t,t')}) into Eq.(\ref{eq:Sigm(m,m')}) leads to an incorrect result, since the terms consisting of the noise spectra only have no explicit time dependence. To derive a correct result, one must go back to the ensemble average (\ref{eq:ensemble_SS'}). 
Here, to show the essence, we only demonstrate the calculation in the following case:  
\begin{eqnarray}
&&I(t,t') =
\int_{-\infty}^{\infty}df_1 \,
df_2 \,
df_3 \,
df_4 \,
df_5 \,
df_6 \,\,
\widetilde{Q}(f_3)\,\widetilde{Q}(f_6)\,
\,e^{i\,2\pi\{(f_1-f_2)t+(f_4-f_5)t'\}}
\nonumber \\
&& \quad\quad\quad\quad~~~~\times\,
\delta_{\tau}(f_1-f_3)\delta_{\tau}(f_3-f_2)
\delta_{\tau}(f_4-f_6)\delta_{\tau}(f_6-f_5)\,
\bigl\langle 
\widetilde{n}_{\scriptscriptstyle I}^*(f_1)
\widetilde{n}_{\scriptscriptstyle I}^*(f_4)
\bigr\rangle\,\bigl\langle 
\widetilde{n}_{\scriptscriptstyle J}(f_2)
\widetilde{n}_{\scriptscriptstyle J}(f_5)
\bigr\rangle,    
\nonumber
\end{eqnarray}
which gives a non-vanishing contribution to 
$\Sigma_{\scriptscriptstyle IJ}^2(t,t')$ [Eq.(\ref{eq:def_Sigma_IJ})]. 
Substituting it into Eq.(\ref{eq:Sigm(m,m')}), with a help of definition (\ref{eq:def_noise}), one obtains 
\begin{eqnarray}
I_{mm'}&=&\frac{1}{T_{\rm orbit}^2}\,\int_0^{T_{\rm orbit}}dt 
\int_0^{T_{\rm orbit}}dt'\,\,I(t,t')\,e^{i\,\omega\,(mt-m't')}
\nonumber\\
&=&\frac{(-1)^{m-m'}}{4\,T_{\rm orbit}^2}\,
\int_{-\infty}^{\infty}df_1 \,
df_2 \,
df_3 \,
df_6 \,\,
\widetilde{Q}(f_3)\,\widetilde{Q}(f_6)\,
\delta_{T_{\rm orbit}}\left(f_1+f_2+\frac{m\omega}{2\pi}\right)
\delta_{T_{\rm orbit}}\left(f_1+f_2+\frac{m'\omega}{2\pi}\right)
\nonumber \\
&& \quad\quad\quad\quad~~~~\times\,
\delta_{\tau}(f_1-f_3)\delta_{\tau}(f_3+f_2)
\delta_{\tau}(f_1+f_6)\delta_{\tau}(f_6-f_2)\,\,
\,N_{\scriptscriptstyle I}(f_1)N_{\scriptscriptstyle J}(f_2), 
\nonumber
\end{eqnarray}
where we used the fact that (\ref{eq:def_delta_tau}) and 
$\omega=2\pi/T_{\rm orbit}$. Since the local observation time $\tau$ is much shorter than the orbital period $T_{\rm orbit}$, the effective support of $\delta_{T_{\rm orbit}}$ is very narrow in frequency space compared with the effective support of $\delta_{\tau}$. 
Hence, one may regard $\delta_{T_{\rm orbit}}$ as ordinary Dirac delta function. This gives 
\begin{eqnarray}
I_{mm'}&=& \frac{(-1)^{m-m'}}{4\,T_{\rm orbit}^2}\,\,
\delta_{T_{\rm orbit}}\left(\frac{(m'-m)\omega}{2\pi}\right)\,
\int_{-\infty}^{\infty}df_1 \,
df_3 \,
df_6 \,\,
\widetilde{Q}(f_3)\,\widetilde{Q}(f_6)\,
\nonumber \\
&& \quad\quad\quad~~~~\times\,
\delta_{\tau}(f_1-f_3)
\delta_{\tau}\left(f_3-f_1-\frac{m\omega}{2\pi}\right)
\delta_{\tau}(f_1+f_6)
\delta_{\tau}\left(f_6+f_1+\frac{m\omega}{2\pi}\right)\,\,
\,N_{\scriptscriptstyle I}(f_1)
N_{\scriptscriptstyle J}\left(f_1+\frac{m\omega}{2\pi}\right). 
\nonumber
\end{eqnarray}
In the above expression, the quantity $\delta_{T_{\rm orbit}}((m'-m)\omega/2\pi)$ reduces to $\delta_{mm'}$. 
Further, we notice that for $|m|<T_{\rm orbit}/\tau$, the shifting of the arguments by $m\omega$ is negligible for our interest of frequency range. Accordingly we obtain
\begin{eqnarray}
I_{mm'}&\simeq& \delta_{mm'}\,\,
\frac{\tau^2}{2T_{\rm orbit}} \int_{-\infty}^{\infty}
\frac{df}{2}\,\left|\widetilde{Q}(f)\right|^2\,
N_{\scriptscriptstyle I}(f)N_{\scriptscriptstyle J}(f), 
\end{eqnarray}
which appears in the last term of Eq.(\ref{eq:result_Sigm(m,m')}).

\bibliographystyle{apsrev}


\begin{thebibliography}{41}
\expandafter\ifx\csname natexlab\endcsname\relax\def\natexlab#1{#1}\fi
\expandafter\ifx\csname bibnamefont\endcsname\relax
  \def\bibnamefont#1{#1}\fi
\expandafter\ifx\csname bibfnamefont\endcsname\relax
  \def\bibfnamefont#1{#1}\fi
\expandafter\ifx\csname citenamefont\endcsname\relax
  \def\citenamefont#1{#1}\fi
\expandafter\ifx\csname url\endcsname\relax
  \def\url#1{\texttt{#1}}\fi
\expandafter\ifx\csname urlprefix\endcsname\relax\def\urlprefix{URL }\fi
\providecommand{\bibinfo}[2]{#2}
\providecommand{\eprint}[2][]{\url{#2}}

\bibitem[{\citenamefont{Ando et~al.}(2001)}]{Ando:2001ej}
\bibinfo{author}{\bibfnamefont{M.}~\bibnamefont{Ando}} \bibnamefont{et~al.}
  (\bibinfo{collaboration}{TAMA}), \bibinfo{journal}{Phys. Rev. Lett.}
  \textbf{\bibinfo{volume}{86}}, \bibinfo{pages}{3950} (\bibinfo{year}{2001}),
  \eprint{astro-ph/0105473}.

\bibitem[{\citenamefont{Abbott}(2005)}]{Abbott:2005pe}
\bibinfo{author}{\bibfnamefont{B.}~\bibnamefont{Abbott}}
  (\bibinfo{collaboration}{LIGO Scientific}) (\bibinfo{year}{2005}),
  \eprint{gr-qc/0505041}.

\bibitem[{\citenamefont{Abbott et~al.}(2004{\natexlab{a}})}]{Abbott:2003pj}
\bibinfo{author}{\bibfnamefont{B.}~\bibnamefont{Abbott}} \bibnamefont{et~al.}
  (\bibinfo{collaboration}{LIGO Scientific}), \bibinfo{journal}{Phys. Rev.}
  \textbf{\bibinfo{volume}{D69}}, \bibinfo{pages}{122001}
  (\bibinfo{year}{2004}{\natexlab{a}}), \eprint{gr-qc/0308069}.

\bibitem[{\citenamefont{{LIGO Scientific
  Collaboration}}(2005)}]{unknown:2005pu}
\bibinfo{author}{\bibnamefont{{LIGO Scientific Collaboration}}}
  (\bibinfo{collaboration}{LIGO Scientific}) (\bibinfo{year}{2005}),
  \eprint{gr-qc/0508065}.

\bibitem[{\citenamefont{Abbott et~al.}(2005)}]{Abbott:2005ez}
\bibinfo{author}{\bibfnamefont{B.}~\bibnamefont{Abbott}} \bibnamefont{et~al.}
  (\bibinfo{collaboration}{LIGO}) (\bibinfo{year}{2005}),
  \eprint{astro-ph/0507254}.

\bibitem[{\citenamefont{Ando et~al.}(2005)}]{Ando:2004rr}
\bibinfo{author}{\bibfnamefont{M.}~\bibnamefont{Ando}} \bibnamefont{et~al.},
  \bibinfo{journal}{Phys. Rev.} \textbf{\bibinfo{volume}{D71}},
  \bibinfo{pages}{082002} (\bibinfo{year}{2005}), \eprint{gr-qc/0411027}.

\bibitem[{\citenamefont{Abbott et~al.}(2004{\natexlab{b}})}]{Abbott:2004rt}
\bibinfo{author}{\bibfnamefont{B.}~\bibnamefont{Abbott}} \bibnamefont{et~al.}
  (\bibinfo{collaboration}{LIGO}), \bibinfo{journal}{Phys. Rev.}
  \textbf{\bibinfo{volume}{D69}}, \bibinfo{pages}{102001}
  (\bibinfo{year}{2004}{\natexlab{b}}).

\bibitem[{\citenamefont{Enoki et~al.}(2004)\citenamefont{Enoki, Inoue,
  Nagashima, and Sugiyama}}]{Enoki:2004ew}
\bibinfo{author}{\bibfnamefont{M.}~\bibnamefont{Enoki}},
  \bibinfo{author}{\bibfnamefont{K.~T.} \bibnamefont{Inoue}},
  \bibinfo{author}{\bibfnamefont{M.}~\bibnamefont{Nagashima}},
  \bibnamefont{and} \bibinfo{author}{\bibfnamefont{N.}~\bibnamefont{Sugiyama}},
  \bibinfo{journal}{Astrophys. J.} \textbf{\bibinfo{volume}{615}},
  \bibinfo{pages}{19} (\bibinfo{year}{2004}), \eprint{astro-ph/0404389}.

\bibitem[{\citenamefont{Koushiappas and Zentner}(2005)}]{Koushiappas:2005qz}
\bibinfo{author}{\bibfnamefont{S.~M.} \bibnamefont{Koushiappas}}
  \bibnamefont{and} \bibinfo{author}{\bibfnamefont{A.~R.}
  \bibnamefont{Zentner}} (\bibinfo{year}{2005}), \eprint{astro-ph/0503511}.

\bibitem[{\citenamefont{Sesana et~al.}(2004)\citenamefont{Sesana, Haardt,
  Madau, and Volonteri}}]{Sesana:2004sp}
\bibinfo{author}{\bibfnamefont{A.}~\bibnamefont{Sesana}},
  \bibinfo{author}{\bibfnamefont{F.}~\bibnamefont{Haardt}},
  \bibinfo{author}{\bibfnamefont{P.}~\bibnamefont{Madau}}, \bibnamefont{and}
  \bibinfo{author}{\bibfnamefont{M.}~\bibnamefont{Volonteri}},
  \bibinfo{journal}{Astrophys. J.} \textbf{\bibinfo{volume}{611}},
  \bibinfo{pages}{623} (\bibinfo{year}{2004}), \eprint{astro-ph/0401543}.

\bibitem[{\citenamefont{Sesana et~al.}(2005)\citenamefont{Sesana, Haardt,
  Madau, and Volonteri}}]{Sesana:2004gf}
\bibinfo{author}{\bibfnamefont{A.}~\bibnamefont{Sesana}},
  \bibinfo{author}{\bibfnamefont{F.}~\bibnamefont{Haardt}},
  \bibinfo{author}{\bibfnamefont{P.}~\bibnamefont{Madau}}, \bibnamefont{and}
  \bibinfo{author}{\bibfnamefont{M.}~\bibnamefont{Volonteri}},
  \bibinfo{journal}{Astrophys. J.} \textbf{\bibinfo{volume}{623}},
  \bibinfo{pages}{23} (\bibinfo{year}{2005}), \eprint{astro-ph/0409255}.

\bibitem[{\citenamefont{Smith et~al.}(2005)\citenamefont{Smith, Kamionkowski,
  and Cooray}}]{Smith:2005mm}
\bibinfo{author}{\bibfnamefont{T.~L.} \bibnamefont{Smith}},
  \bibinfo{author}{\bibfnamefont{M.}~\bibnamefont{Kamionkowski}},
  \bibnamefont{and} \bibinfo{author}{\bibfnamefont{A.}~\bibnamefont{Cooray}}
  (\bibinfo{year}{2005}), \eprint{astro-ph/0506422}.

\bibitem[{\citenamefont{Wyithe and Loeb}(2003)}]{Wyithe:2002ep}
\bibinfo{author}{\bibfnamefont{J.~S.~B.} \bibnamefont{Wyithe}}
  \bibnamefont{and} \bibinfo{author}{\bibfnamefont{A.}~\bibnamefont{Loeb}},
  \bibinfo{journal}{Astrophys. J.} \textbf{\bibinfo{volume}{590}},
  \bibinfo{pages}{691} (\bibinfo{year}{2003}), \eprint{astro-ph/0211556}.

\bibitem[{\citenamefont{Bender et~al.}(1998)}]{Bender:1998}
\bibinfo{author}{\bibfnamefont{P.~L.} \bibnamefont{Bender}}
  \bibnamefont{et~al.}, \bibinfo{journal}{LISA Pre-Phase A Report}
  (\bibinfo{year}{1998}).

\bibitem[{\citenamefont{Seto et~al.}(2001)\citenamefont{Seto, Kawamura, and
  Nakamura}}]{Seto:2001qf}
\bibinfo{author}{\bibfnamefont{N.}~\bibnamefont{Seto}},
  \bibinfo{author}{\bibfnamefont{S.}~\bibnamefont{Kawamura}}, \bibnamefont{and}
  \bibinfo{author}{\bibfnamefont{T.}~\bibnamefont{Nakamura}},
  \bibinfo{journal}{Phys. Rev. Lett.} \textbf{\bibinfo{volume}{87}},
  \bibinfo{pages}{221103} (\bibinfo{year}{2001}), \eprint{astro-ph/0108011}.

\bibitem[{\citenamefont{Phinney et~al.}(2003)}]{PhinneyBBO:2003}
\bibinfo{author}{\bibfnamefont{E.~S.} \bibnamefont{Phinney}}
  \bibnamefont{et~al.}, \bibinfo{journal}{NASA Mission Concept Study}
  (\bibinfo{year}{2003}).

\bibitem[{BBO(2003)}]{BBO:2003}
\bibinfo{journal}{URL http://universe.nasa.gov/be/roadmap.html}
  (\bibinfo{year}{2003}).

\bibitem[{\citenamefont{Kawamura et~al.}(2005)}]{FP-DECIGO:2005}
\bibinfo{author}{\bibfnamefont{S.}~\bibnamefont{Kawamura}}
  \bibnamefont{et~al.}, \bibinfo{journal}{Amaldi 6 Conference, Japan}
  (\bibinfo{year}{2005}).

\bibitem[{\citenamefont{Kudoh and Taruya}(2005)}]{Kudoh:2004he}
\bibinfo{author}{\bibfnamefont{H.}~\bibnamefont{Kudoh}} \bibnamefont{and}
  \bibinfo{author}{\bibfnamefont{A.}~\bibnamefont{Taruya}},
  \bibinfo{journal}{Phys. Rev.} \textbf{\bibinfo{volume}{D71}},
  \bibinfo{pages}{024025} (\bibinfo{year}{2005}), \eprint{gr-qc/0411017}.

\bibitem[{\citenamefont{Taruya and Kudoh}(2005)}]{Taruya:2005yf}
\bibinfo{author}{\bibfnamefont{A.}~\bibnamefont{Taruya}} \bibnamefont{and}
  \bibinfo{author}{\bibfnamefont{H.}~\bibnamefont{Kudoh}},
  \bibinfo{journal}{Phys. Rev.} \textbf{\bibinfo{volume}{D72}},
  \bibinfo{pages}{104015} (\bibinfo{year}{2005}), \eprint{gr-qc/0507114}.

\bibitem[{\citenamefont{Flanagan}(1993)}]{Flanagan:1993ix}
\bibinfo{author}{\bibfnamefont{E.~E.} \bibnamefont{Flanagan}},
  \bibinfo{journal}{Phys. Rev.} \textbf{\bibinfo{volume}{D48}},
  \bibinfo{pages}{2389} (\bibinfo{year}{1993}), \eprint{astro-ph/9305029}.

\bibitem[{\citenamefont{Allen and Romano}(1999)}]{Allen:1997ad}
\bibinfo{author}{\bibfnamefont{B.}~\bibnamefont{Allen}} \bibnamefont{and}
  \bibinfo{author}{\bibfnamefont{J.~D.} \bibnamefont{Romano}},
  \bibinfo{journal}{Phys. Rev.} \textbf{\bibinfo{volume}{D59}},
  \bibinfo{pages}{102001} (\bibinfo{year}{1999}), \eprint{gr-qc/9710117}.

\bibitem[{\citenamefont{Shaddock et~al.}(2003)\citenamefont{Shaddock, Tinto,
  Estabrook, and Armstrong}}]{Shaddock:2003dj}
\bibinfo{author}{\bibfnamefont{D.~A.} \bibnamefont{Shaddock}},
  \bibinfo{author}{\bibfnamefont{M.}~\bibnamefont{Tinto}},
  \bibinfo{author}{\bibfnamefont{F.~B.} \bibnamefont{Estabrook}},
  \bibnamefont{and} \bibinfo{author}{\bibfnamefont{J.~W.}
  \bibnamefont{Armstrong}}, \bibinfo{journal}{Phys. Rev.}
  \textbf{\bibinfo{volume}{D68}}, \bibinfo{pages}{061303}
  (\bibinfo{year}{2003}), \eprint{gr-qc/0307080}.

\bibitem[{\citenamefont{Tinto et~al.}(2004)\citenamefont{Tinto, Estabrook, and
  Armstrong}}]{Tinto:2003vj}
\bibinfo{author}{\bibfnamefont{M.}~\bibnamefont{Tinto}},
  \bibinfo{author}{\bibfnamefont{F.~B.} \bibnamefont{Estabrook}},
  \bibnamefont{and} \bibinfo{author}{\bibfnamefont{J.~W.}
  \bibnamefont{Armstrong}}, \bibinfo{journal}{Phys. Rev.}
  \textbf{\bibinfo{volume}{D69}}, \bibinfo{pages}{082001}
  (\bibinfo{year}{2004}), \eprint{gr-qc/0310017}.

\bibitem[{\citenamefont{Allen and Ottewill}(1997)}]{Allen:1997gp}
\bibinfo{author}{\bibfnamefont{B.}~\bibnamefont{Allen}} \bibnamefont{and}
  \bibinfo{author}{\bibfnamefont{A.~C.} \bibnamefont{Ottewill}},
  \bibinfo{journal}{Phys. Rev.} \textbf{\bibinfo{volume}{D56}},
  \bibinfo{pages}{545} (\bibinfo{year}{1997}), \eprint{gr-qc/9607068}.

\bibitem[{\citenamefont{Armstrong et~al.}(1999)\citenamefont{Armstrong,
  Estabrook, and Tinto}}]{Armstrong:1999}
\bibinfo{author}{\bibfnamefont{J.~W.} \bibnamefont{Armstrong}},
  \bibinfo{author}{\bibfnamefont{F.~B.} \bibnamefont{Estabrook}},
  \bibnamefont{and} \bibinfo{author}{\bibfnamefont{M.}~\bibnamefont{Tinto}},
  \bibinfo{journal}{Astrophys. J.} \textbf{\bibinfo{volume}{527}},
  \bibinfo{pages}{814} (\bibinfo{year}{1999}).

\bibitem[{\citenamefont{Larson et~al.}(2000)\citenamefont{Larson, Hiscock, and
  Hellings}}]{Larson:1999we}
\bibinfo{author}{\bibfnamefont{S.~L.} \bibnamefont{Larson}},
  \bibinfo{author}{\bibfnamefont{W.~A.} \bibnamefont{Hiscock}},
  \bibnamefont{and} \bibinfo{author}{\bibfnamefont{R.~W.}
  \bibnamefont{Hellings}}, \bibinfo{journal}{Phys. Rev.}
  \textbf{\bibinfo{volume}{D62}}, \bibinfo{pages}{062001}
  (\bibinfo{year}{2000}), \eprint{gr-qc/9909080}.

\bibitem[{\citenamefont{Schilling}(1997)}]{Schilling:1997id}
\bibinfo{author}{\bibfnamefont{R.}~\bibnamefont{Schilling}},
  \bibinfo{journal}{Class. Quant. Grav.} \textbf{\bibinfo{volume}{14}},
  \bibinfo{pages}{1513} (\bibinfo{year}{1997}).

\bibitem[{\citenamefont{Cornish}(2002{\natexlab{a}})}]{Cornish:2001bb}
\bibinfo{author}{\bibfnamefont{N.~J.} \bibnamefont{Cornish}},
  \bibinfo{journal}{Phys. Rev.} \textbf{\bibinfo{volume}{D65}},
  \bibinfo{pages}{022004} (\bibinfo{year}{2002}{\natexlab{a}}),
  \eprint{gr-qc/0106058}.

\bibitem[{\citenamefont{Schneider et~al.}(2001)\citenamefont{Schneider,
  Ferrari, Matarrese, and Portegies~Zwart}}]{Schneider:2000sg}
\bibinfo{author}{\bibfnamefont{R.}~\bibnamefont{Schneider}},
  \bibinfo{author}{\bibfnamefont{V.}~\bibnamefont{Ferrari}},
  \bibinfo{author}{\bibfnamefont{S.}~\bibnamefont{Matarrese}},
  \bibnamefont{and} \bibinfo{author}{\bibfnamefont{S.~F.}
  \bibnamefont{Portegies~Zwart}}, \bibinfo{journal}{Mon. Not. Roy. Astron.
  Soc.} \textbf{\bibinfo{volume}{324}}, \bibinfo{pages}{797}
  (\bibinfo{year}{2001}), \eprint{astro-ph/0002055}.

\bibitem[{\citenamefont{Farmer and Phinney}(2003)}]{Farmer:2003pa}
\bibinfo{author}{\bibfnamefont{A.~J.} \bibnamefont{Farmer}} \bibnamefont{and}
  \bibinfo{author}{\bibfnamefont{E.~S.} \bibnamefont{Phinney}},
  \bibinfo{journal}{Mon. Not. Roy. Astron. Soc.}
  \textbf{\bibinfo{volume}{346}}, \bibinfo{pages}{1197} (\bibinfo{year}{2003}),
  \eprint{astro-ph/0304393}.

\bibitem[{\citenamefont{Seto}(2005)}]{Seto:2005qy}
\bibinfo{author}{\bibfnamefont{N.}~\bibnamefont{Seto}} (\bibinfo{year}{2005}),
  \eprint{gr-qc/0510067}.

\bibitem[{\citenamefont{Cutler and Harms}(2005)}]{Cutler:2005qq}
\bibinfo{author}{\bibfnamefont{C.}~\bibnamefont{Cutler}} \bibnamefont{and}
  \bibinfo{author}{\bibfnamefont{J.}~\bibnamefont{Harms}}
  (\bibinfo{year}{2005}), \eprint{gr-qc/0511092}.

\bibitem[{\citenamefont{Buonanno et~al.}(2005)\citenamefont{Buonanno, Sigl,
  Raffelt, Janka, and Muller}}]{Buonanno:2004tp}
\bibinfo{author}{\bibfnamefont{A.}~\bibnamefont{Buonanno}},
  \bibinfo{author}{\bibfnamefont{G.}~\bibnamefont{Sigl}},
  \bibinfo{author}{\bibfnamefont{G.~G.} \bibnamefont{Raffelt}},
  \bibinfo{author}{\bibfnamefont{H.-T.} \bibnamefont{Janka}}, \bibnamefont{and}
  \bibinfo{author}{\bibfnamefont{E.}~\bibnamefont{Muller}},
  \bibinfo{journal}{Phys. Rev.} \textbf{\bibinfo{volume}{D72}},
  \bibinfo{pages}{084001} (\bibinfo{year}{2005}), \eprint{astro-ph/0412277}.

\bibitem[{\citenamefont{Bennett et~al.}(1996)}]{Bennett:1996ce}
\bibinfo{author}{\bibfnamefont{C.~L.} \bibnamefont{Bennett}}
  \bibnamefont{et~al.}, \bibinfo{journal}{Astrophys. J.}
  \textbf{\bibinfo{volume}{464}}, \bibinfo{pages}{L1} (\bibinfo{year}{1996}),
  \eprint{astro-ph/9601067}.

\bibitem[{\citenamefont{Bennett et~al.}(2003)}]{Bennett:2003bz}
\bibinfo{author}{\bibfnamefont{C.~L.} \bibnamefont{Bennett}}
  \bibnamefont{et~al.}, \bibinfo{journal}{Astrophys. J. Suppl.}
  \textbf{\bibinfo{volume}{148}}, \bibinfo{pages}{1} (\bibinfo{year}{2003}),
  \eprint{astro-ph/0302207}.

\bibitem[{\citenamefont{Cornish}(2002{\natexlab{b}})}]{Cornish:2002bh}
\bibinfo{author}{\bibfnamefont{N.~J.} \bibnamefont{Cornish}},
  \bibinfo{journal}{Class. Quant. Grav.} \textbf{\bibinfo{volume}{19}},
  \bibinfo{pages}{1279} (\bibinfo{year}{2002}{\natexlab{b}}).

\bibitem[{\citenamefont{Seto and Cooray}(2004)}]{Seto:2004np}
\bibinfo{author}{\bibfnamefont{N.}~\bibnamefont{Seto}} \bibnamefont{and}
  \bibinfo{author}{\bibfnamefont{A.}~\bibnamefont{Cooray}}
  (\bibinfo{year}{2004}), \eprint{astro-ph/0403259}.

\bibitem[{\citenamefont{Blake and Wall}(2002)}]{Blake:2002gx}
\bibinfo{author}{\bibfnamefont{C.}~\bibnamefont{Blake}} \bibnamefont{and}
  \bibinfo{author}{\bibfnamefont{J.}~\bibnamefont{Wall}},
  \bibinfo{journal}{Nature} \textbf{\bibinfo{volume}{416}},
  \bibinfo{pages}{150} (\bibinfo{year}{2002}), \eprint{astro-ph/0203385}.

\bibitem[{\citenamefont{Cornish}(2001)}]{Cornish:2001hg}
\bibinfo{author}{\bibfnamefont{N.~J.} \bibnamefont{Cornish}},
  \bibinfo{journal}{Class. Quant. Grav.} \textbf{\bibinfo{volume}{18}},
  \bibinfo{pages}{4277} (\bibinfo{year}{2001}), \eprint{astro-ph/0105374}.

\bibitem[{\citenamefont{Prince et~al.}(2002)\citenamefont{Prince, Tinto,
  Larson, and Armstrong}}]{Prince:2002hp}
\bibinfo{author}{\bibfnamefont{T.~A.} \bibnamefont{Prince}},
  \bibinfo{author}{\bibfnamefont{M.}~\bibnamefont{Tinto}},
  \bibinfo{author}{\bibfnamefont{S.~L.} \bibnamefont{Larson}},
  \bibnamefont{and} \bibinfo{author}{\bibfnamefont{J.~W.}
  \bibnamefont{Armstrong}}, \bibinfo{journal}{Phys. Rev.}
  \textbf{\bibinfo{volume}{D66}}, \bibinfo{pages}{122002}
  (\bibinfo{year}{2002}), \eprint{gr-qc/0209039}.

\end{thebibliography}

\end{document}